\documentclass[sigconf]{acmart}
\usepackage{amsmath,amsfonts}
\usepackage{wasysym}
\usepackage{dsfont}
\usepackage{algorithm}

\usepackage[noend]{algpseudocode}

\usepackage{array}
\usepackage{makecell}
\usepackage[labelfont=scriptsize,textfont=scriptsize]{subfig}
\usepackage{textcomp}

\usepackage{stfloats}
\usepackage{tcolorbox}
\usepackage{xurl}
\usepackage{multirow}
\usepackage{threeparttable}
\usepackage{booktabs}
\usepackage{multirow}
\usepackage{verbatim}
\usepackage{color, soul}
\usepackage{graphicx}
\usepackage{xcolor}
\usepackage{xspace}
\usepackage{threeparttable}
\newcommand{\sys}{\textsc{PhishEye}\xspace}
\usepackage{tabularx}

\usepackage{xurl}

\newcommand{\graphname}{HTAMG\xspace}
\usepackage{hyperref}
\hypersetup{colorlinks=true, citecolor= blue,linkcolor=blue, filecolor=blue,urlcolor=blue}
	
\newenvironment{inditemize}{
\begin{list}{$\bullet$}{
\setlength{\labelwidth}{-6pt}
\setlength{\itemsep}{0pt}
\setlength{\leftmargin}{\labelwidth}
\addtolength{\leftmargin}{\labelsep}
\setlength{\parindent}{0pt}
\setlength{\listparindent}{\parindent}
\setlength{\parsep}{0pt}
\setlength{\topsep}{0pt}}}{\end{list}}

\makeatletter
\def\hlinewd#1{%
	\noalign{\ifnum0=`}\fi\hrule \@height #1 %
	\futurelet\reserved@a\@xhline}
\makeatother

\setcopyright{cc}
\setcctype[4.0]{by-nc-nd}
\copyrightyear{2026}
\acmYear{2026}

\acmConference[CCS '26]{ACM Conference on Computer and Communications Security}{November 15--19, 2026}{Hague, Netherlands}

\makeatletter
\AtBeginEnvironment{figure}{\@Description@presenttrue}
\AtBeginEnvironment{figure*}{\@Description@presenttrue}
\makeatother

\AtBeginDocument{\let\balance\relax}

\begin{document}
\title{Phishing Detection in Ethereum via Temporal \\ Graph Contrastive Learning}

\author{Cong Wu}
\affiliation{%
 \institution{Wuhan University}
  \city{Wuhan}
  \country{China}}
\email{cnacwu@whu.edu.cn}

\author{Jing Chen}
\affiliation{%
 \institution{Wuhan University}
  \city{Wuhan}
  \country{China}}
\email{chenjing@whu.edu.cn}

\author{Siqi Lin}
\affiliation{%
 \institution{Wuhan University}
  \city{Wuhan}
  \country{China}}
\email{linsiqi@whu.edu.cn}

\author{Hongda Li}
\affiliation{%
 \institution{Palo Alto Networks}
  \city{California}
  \country{USA}}
\email{hongdaljz@gmail.com}

\author{Ziming Zhao}
\affiliation{%
 \institution{Northeastern University}
  \city{Boston}
  \country{USA}}
\email{z.zhao@northeastern.edu}

\renewcommand{\shortauthors}{Cong Wu et al.}
\renewcommand{\shorttitle}{Phishing Detection in Ethereum via Temporal Graph Contrastive Learning}

\begin{abstract}
	Blockchain and decentralized finance have revolutionized the financial ecosystem while simultaneously exposing it to cryptocurrency phishing attacks. 
	Existing phishing detection methods primarily rely on graph learning, but they face significant limitations. Static graph learning approaches fail to account for the temporal evolution of phishing patterns, while semi-dynamic methods, such as those combining static GNNs with LSTM, struggle to capture the irregular and bursty nature of blockchain transactions.
	Moreover, these methods overlook the diversity of Ethereum transactions, treating them as homogeneous graphs, and heavily rely on supervised learning, which requires extensive labeled data that is not readily available. These limitations reduce their adaptability to emerging phishing threats.

	In this paper, we present \sys, a fully dynamic self-supervised system that monitors on-chain transactions to detect phishing activities. \sys formulates Ethereum transactions as a heterogeneous temporal attributed multi-graph and incorporates a novel temporal graph contrastive learning model, which captures both temporal patterns and heterogeneous transaction types. 
	The evaluation on a dataset of 161,658 addresses and 416,541 transactions shows that \sys outperforms existing methods, achieving an F1 score of 87.23\% and an AUC of 98.43\% for phishing transaction detection, and an F1 score of 94.19\% and an AUC of 98.03\% for phishing account detection. 
	In real-world deployment from May 1, 2023 to July 31, 2024, \sys identified 1,803 previously unknown phishing addresses, providing early alerts that helped prevent losses exceeding 2 billion USD.

\end{abstract}

\ccsdesc[500]{Security and privacy~Distributed systems security}

\keywords{Phishing detection, Ethereum, temporal graph learning, contrastive learning}

\maketitle

\section{Introduction}

Blockchain and smart contracts are reshaping the financial landscape and have opened the door to decentralized finance (DeFi), which reached a total value locked of \$46.18B on Ethereum by April 2026~\cite{TVL,zhou2023sok,zhou2021high}.
Yet, this rapid growth has simultaneously made DeFi a prime target for cryptocurrency phishing attacks~\cite{chainalysis,liu2024fishing,guan2024characterizing,he2025phishing,ghosh2023investigating}.
Such attacks are a form of social engineering, where adversaries deceive victims using tactics like fake bounties or airdrops involving crypto tokens~\cite{forta_phishing,forbes,li2022security,scharfman2024crypto,varshney2024anti}.
Attackers often impersonate legitimate entities to deceive users into repeatedly transferring funds to malicious addresses within a short time frame,
transferring funds to the malicious phishing address, then quickly move the stolen funds to other accounts to evade detection.
In 2024 alone, phishing scams led to over \$1B across 296 incidents~\cite{cointelegraph}. These attacks exploit the permissionless and anonymous nature of blockchain, making detection and prevention both technically challenging and increasingly urgent.

Ethereum transactions naturally form a graph, with accounts as nodes and transactions as edges, which enables phishing detection to be framed as a graph learning task~\cite{liu2025phishing,xiao2025pheromone}. Early approaches, known as static graph learning methods like Trans2vec~\cite{wu2020phishers}, apply graph neural networks (GNNs) to capture relationships between accounts~\cite{ghosh2025catalog,alghuried2026phishing,fu2024ct,huang2024peae}. However, by treating the graph structure as fixed, these methods fail to capture the temporal evolution of phishing patterns and cannot adapt to the continual emergence of new accounts and transactions. Consequently, static GNNs often miss critical real-time shifts in phishing tactics, limiting their effectiveness in detecting emerging phishing campaigns~\cite{liu2023graphsage}.

Some methods, such as TTAGN~\cite{li2022ttagn} and TGC~\cite{li2023tgc}, extend static graph learning by combining GNNs with temporal models, such as Long Short-Term Memory (LSTM), to capture sequential changes over time~\cite{huang2023ethereum,li2023siege,lv2023phishing}.
While this adds a temporal component, such models including LSTM assume that events occur at fixed or uniformly spaced intervals, overlooking the irregular timing and bursty nature of blockchain transactions. For this reason, these approaches are often referred to as semi-dynamic: they incorporate temporal dynamics only partially, without fully adapting to the highly dynamic and unpredictable nature of phishing behaviors. As a result, they frequently miss critical real-time shifts in phishing tactics, limiting their effectiveness in detecting time-sensitive phishing campaigns.

Beyond the aforementioned limitations, existing static and semi-dynamic approaches all rely on supervised learning, requiring large volumes of human-labeled phishing data that are not readily available. 
In addition, two fundamental but often overlooked aspects of Ethereum further hinders effective graph learning: 
(i) since Ethereum accounts provide little descriptive information, it is difficult to construct meaningful transaction graphs, which weakens the model's ability to capture essential patterns and reduces detection accuracy~\cite{wu2024tokenscout}; 
and (ii) Ethereum transactions are also highly diverse, involving different types and account behaviors, yet most methods treat them as homogeneous graphs. 
For example, fungible token (FT) and non-fungible token (NFT) transactions differ substantially. 
NFTs include unique token IDs and specialized smart contract interactions, while FTs are interchangeable and typically do not involve such unique identifiers or interactions~\cite{ma2025sok,peelam2025decentralized,lavrova2022nft,jy2024nft}. Overlooking such heterogeneity oversimplifies the transaction graph and limits effectiveness in detecting phishing across different transaction types.

In this paper, we present \sys, a fully dynamic self-supervised system that monitors the on-chain transaction graph to detect phishing transactions and addresses.
\sys formulates the temporal DeFi transactions as a \underline{h}eterogeneous \underline{t}emporal \underline{a}ttributed \underline{m}ulti-\underline{g}raph (\graphname) by extracting comprehensive node attributes from the transactional activity and network structure.
\sys incorporates a novel temporal graph contrastive learning model for phishing (PhishTGL) that encodes time as a learnable representation with temporal attention, node memory, and temporal message passing, going beyond semi-dynamic approaches that simply combine LSTM with static GNNs~\cite{li2022ttagn,li2023tgc}. This design enables accurate modeling of temporal transaction patterns while the node memory mechanism preserves historical transaction dynamics.
To comprehensively aggregate neighbor representations, \sys employs a temporal attention mechanism that dynamically weighs the importance of neighboring nodes based on their temporal interactions, effectively capturing the evolving nature of transaction activities. In addition, PhishTGL is designed with a self-supervised graph learning strategy via node-level graph contrastive learning, which enables robust representation learning across diverse graph views and imbalanced datasets.

To evaluate \sys, we compiled a phishing dataset containing 161,658 addresses and 416,541 transactions. The dataset was sourced from two anonymized companies, with phishing labels verified through expert analysis and real-world victim disclosures. On this dataset, \sys achieved an F1 score of 87.23\% and an AUC of 98.43\% for phishing transaction detection, and demonstrated robustness in phishing account detection with an F1 score of 94.19\% and an AUC of 98.03\%.
Additionally, we deployed \sys on the Ethereum network with an anonymized company to enable real-time monitoring and tracking of phishing activities. By integrating directly with the blockchain, \sys provides timely alerts of potential phishing incidents. During deployment, the prototype detected 2,153 phishing addresses between May 1, 2023, and July 31, 2024, of which 1,803 were identified before any public reports and later confirmed through victim disclosures and phishing site analysis. Another 196 addresses were flagged as suspicious and validated by risk auditors. In total, \sys helped alert losses of \$2,039.1M out of an estimated \$2,813.7M in phishing-related losses during this period. The detected addresses were responsibly disclosed to two anonymized companies for further alerting and tracking. In this deployment, \sys outperformed existing methods by achieving a false negative rate of 1.31\% and a false positive rate of 2.75\%, significantly better than TTAGN (19.65\%, 26.19\%), TGC (18.72\%, 32.14\%), and Trans2vec (11.37\%, 19.43\%).
The contributions of this paper are as follows:

\begin{itemize}
	\item We present \sys, the first self-supervised system for phishing detection on Ethereum, built on heterogeneous temporal graph contrastive learning and capable of modeling diverse transaction types including Ether, FT, and NFT transfers.

	\item We present PhishTGL, a novel temporal graph learning model that integrates temporal encoding, attention-based aggregation, and contrastive learning to achieve robust phishing detection.

	\item We comprehensively evaluate \sys across diverse settings and real-world deployments, showing consistent improvements over existing phishing detection and graph learning methods, and further analyze the fund flows of newly detected phishing addresses to characterize post-attack behaviors.
	
	\item We contribute a new cryptocurrency phishing dataset for public use, including 1,000 phishing and 160,658 non-phishing addresses, along with 48,135 phishing and 368,406 non-phishing transactions. We open-source the dataset and code of \sys for reproducibility at the repository: \textcolor{red}{\url{https://anonymous.4open.science/r/PhishEye-3CEE/}}.
\end{itemize}

\section{Background}

\textbf{Ethereum and tokens.}
Ethereum is a leading blockchain platform known for its support of smart contracts and a wide variety of tokens~\cite{he2023tokenaware}. 
Beyond cryptocurrency transactions, it enables the use of diverse token types, including fungible tokens (FTs) such as USDT and ERC-20, and non-fungible tokens (NFTs) like ERC-721 and ERC-1155~\cite{chen2019tokenscope,cernera2023token}.
These tokens are pivotal to the Ethereum ecosystem, representing distinct assets or values and enabling varied applications and financial activities~\cite{song2025automated}.
Tokens are managed and interacted with via smart contracts, which autonomously execute and enforce code-based agreements, forming the foundation for diverse financial activities and applications in the web3 ecosystem~\cite{ma2023pied,liang2025vulseye,wu2024tokenscout,zhou2024stop}.

\textbf{Accounts and transactions.}
Ethereum operates through a network of accounts and transactions, forming the backbone of its decentralized finance~\cite{zhao2021temporal}. It features two primary types of accounts: externally owned accounts (EOAs), controlled by private keys, and contract accounts (CAs), governed by embedded smart contract code~\cite{liang2024ponziguard}. Ethereum's transaction system is multifaceted, enabling actions such as Ether transfers, contract interactions, and the deployment of new smart contracts. Once recorded on the blockchain, these transactions are immutable.
Ethereum transactions encompass various types, including basic Ether transfers, the creation of new smart contracts, and interactions with existing contracts. Additionally, internal transactions, triggered by smart contract execution, can transfer Ether or tokens in accordance with contract logic. Together, these diverse transaction types enable Ethereum to function as a global state machine, with its state evolving dynamically with each transaction.

\begin{figure}[!t]
	\centering
\includegraphics[width = 0.98\linewidth]{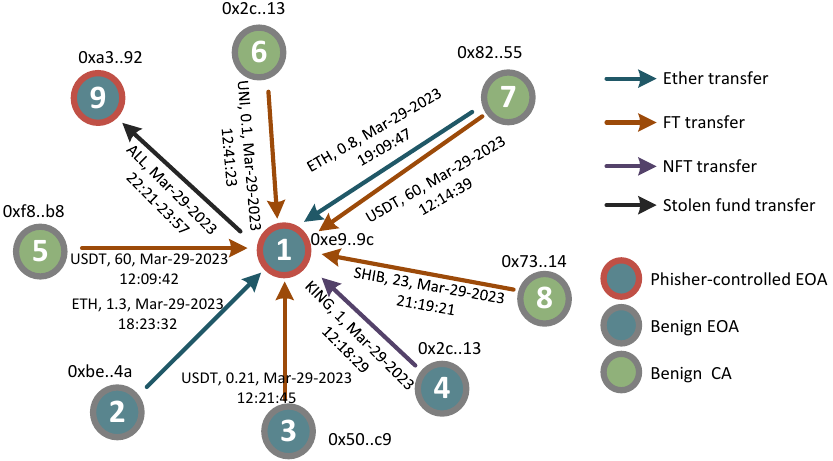}
	\caption{Motivating example of a phishing incident on March 29, 2023} 
	\label{fig:gnn-fig}
\end{figure}

\textbf{A motivating phishing example.}
Figure~\ref{fig:gnn-fig} illustrates a real-world Ethereum phishing incident on March 29, 2023. The phisher-controlled address \texttt{0xe9..9c} (Node 1) received a burst of transfers from multiple victim addresses (Nodes 2-8) within a short interval. 
Finally, the stolen assets were quickly funneled to another address \texttt{0xa3..92} (Node 9) to obscure the trail.
 These inflows included both fungible and non-fungible tokens, indicating that the attackers exploited diverse assets to broaden the attack surface. After receipt, the funds were rapidly moved through additional accounts to obscure provenance and evade detection. This combination of repeated transfers to a single address, multi-asset inflows, and swift laundering reflects a typical phishing pattern and highlights why traditional static methods struggle to capture the full dynamics of such campaigns.

\section{\sys}

In this section, we formalize the blockchain phishing detection problem, present an overview of \sys, and describe its detailed design, including transaction collection, temporal graph construction, temporal graph contrastive learning, and phishing detection.

\subsection{Problem Formulation}

We formulate the complex Ethereum transaction network with a heterogeneous temporal attributed multi-graph (\graphname),
and further conceptualize the Ethereum phishing detection as a graph node and edge classification problem. 
\( HTAMG = (\mathcal{V}, \mathcal{E}, \mathbf{X}^\mathcal{V}, \mathbf{X}^\mathcal{E}, \mathcal{T}, \mathcal{L}) \),
where \( \mathcal{V} \) is a set of nodes representing Ethereum accounts. 
$\mathcal{E}$ $=$ $\{e_1$,$e_2$,$...$$e_k...\}$ represents the temporal transaction edges between accounts. 
Each node $v_i\in \mathcal{V}$ has its original attributes $x_{v_i} \in \mathbf{X}^\mathcal{V}$.
Each temporal transaction edge $ e_k = (v_i,v_j, t_{k})$ represents the transaction from account $v_i$ to $v_j$ at timestamp $t_k$, where $t_k \in \mathcal{T} $, and $v_i$, $v_j\in \mathcal{V}$. 
The attribute of edge $e_k$ is noted as $x_{e_k} \in \mathbf{X}^\mathcal{E}$.
The phishing labels of accounts ($\mathcal{V}$) and transactions ($\mathcal{E}$) are defined as \( \mathcal{L}\in \{0, 1\}^{(|\mathcal{V}|+ |\mathcal{E}|)\times 2} \), which indicate if the accounts and transactions are phisher-controlled.

Notably, it is possible that multiple transactions exist between two different accounts at different timestamps. 
Moreover, the types of each node (account) or edge (transaction) can be different.  
For example, the accounts could be EOA or CA, while the transaction types could be Ether transfers, FT transfers, NFT transfers, contract interactions, or internal contract transactions. 
Therefore, the graph $HTAMG$ is a heterogeneous temporal attributed multi-graph.   
Our goal is to effectively learn how the transactional behaviors of nodes \( \mathcal{V} \) and edges \( \mathcal{E} \) evolve in an Ethereum's dynamic transaction graph, integrating both structural and temporal dynamics to capture time-sensitive features and connectivity patterns.

\subsection{Overview of \sys}
\begin{figure}[!t]
	\centering
	\includegraphics[width = \linewidth]{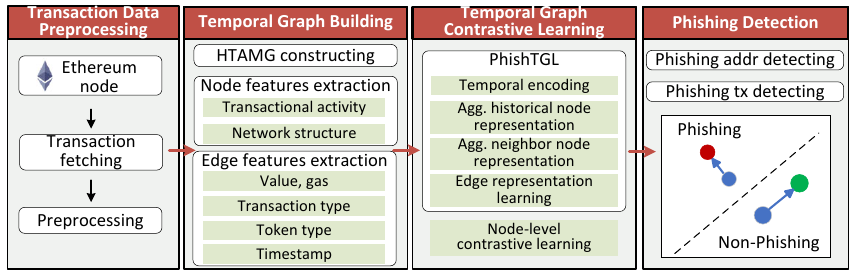}
	\caption{Workflow of \sys}	
	\label{fig:overview}
\end{figure}

As shown in Figure~\ref{fig:overview}, 
\sys consists of four steps:
(i) \emph{Transaction Data Preprocessing}. This step fetches the transaction data from the Ethereum network and preprocesses the data;
(ii) \emph{Temporal Graph Building}. This step structures the preprocessed transactions into \graphname and extracts the node and edge features; 
Unlike previous methods that treat Ethereum transactions as homogeneous graphs, our approach captures the heterogeneity of different transaction types and account behaviors, providing a more accurate and comprehensive representation for phishing detection;
(iii) \emph{Temporal Graph Contrastive Learning}. This step employs PhishTGL, a self-supervised temporal graph model, to learn the representations of nodes and edges in \graphname.
PhishTGL leverages temporal encoding, aggregates historical node representations, aggregates neighbor node representation, and performs edge representation learning.
PhishTGL is trained in node-level contrastive learning without requiring any data label;
(iv) \emph{Phishing Detection}. This step utilizes the representations learned in the PhishTGL to classify nodes and edges as phishing or non-phishing. 

\subsection{Transaction Data Preprocessing}

\sys first parses each transaction and categorizes it into different types: Ether transfers, FT transfers, NFT transfers, internal contract transactions, or contract interactions. \sys excludes zero-value Ether transfers and non-ERC20/721/1155 interactions, which are unrelated to phishing.
For each valid transaction, \sys extracts the key fields, which includes sender and receiver addresses, token type, transfer amount, timestamp, and gas used.
The numerical identifiers are assigned to the transaction and token types to streamline downstream graph processing and learning.

Then \sys preprocesses the fetched data to enhance model training efficiency and performance.
Ethereum accounts, initially in an extended hexadecimal format, are converted into numerical identifiers, starting sequentially from 0, significantly reducing the computational and storage complexity.
Furthermore, transactions are chronologically ordered and timestamps are reformatted numerically in seconds by subtracting the minimum timestamp in the dataset.
\sys also performs z-score normalization on transaction value and gas respectively using the mean \(\mu\) and the standard deviation \(\sigma\) of these fields, i.e., \( z = (x - \mu)/ \sigma \).

\subsection{Temporal Graph Building}
\label{sec:Temporal_Graph_Builder}

\sys then organizes processed transactions into a \graphname, where nodes represent accounts and edges represent transactions with attributes such as token type, amount, and timestamp. The graph is enriched with node and edge features to capture transactional activities and account relationships. 

	\textbf{HTAMG construction.}
HTAMG is created by converting transactions into edges. The senders and recipients are abstracted as nodes. Though the graph evolves over time, the timestamps of transactions are integrated to reflect the dynamic transactional activities. 
The new transaction data forms a testing graph that is aligned with the existing \graphname. This process includes adding edges to existing nodes, creating new nodes and edges for new account transactions, and expanding the graph to accurately reflect the network's current state.

	\textbf{Node features.}
	The node features extraction is vital for efficient graph learning in \graphname, especially provided the lack of inherent attributes in Ethereum accounts. Simple features like historical token balances are not sufficient due to their static nature and limited insight into dynamic interactions.
	While previous methods extract basic features, e.g., in- and out-degree~\cite{li2022ttagn,li2023tgc}, they fail to capture the complexity of account activities and relationships.
\sys overcomes this limitation by identifying a comprehensive feature set indicative of phishing, as shown in Table~\ref{tab:nodefeatures}. The features consist of transactional activity (T1–T9) and network structure (N1–N9), providing a richer and more accurate representation of phishing behavior. Details are given in Appendix~\ref{app:nodefeatures}.

\begin{table*}[ht]
	\centering
	\caption{Extracted transactional activity and network structure features for \graphname.}
	\resizebox{\linewidth}{!}{\begin{tabular}{cll}
		\hline
\textbf{Type} & \textbf{Feature} & \textbf{Description} \\ \hline
\multirow{9}{*}{\rotatebox{90}{Transactional activity (T)}}
& T1. Account indegree (AID) & The number of incoming transactions to the node. AID reflects receiving activity.\\
& T2. Account outdegree (AOD) & The number of outgoing transactions from node. AOD reflects sending activity. \\
& T3. Account total degree (ATD) & Total number of transactions of an account, reflecting overall engagement. \\ 
& T4. Inbound value flow (IVF) & The total received funds, measuring a node's economic interactions within the transaction network. \\ 
& T5. Outbound value flow (OVF) & The total sent funds, measuring a node's economic interactions within the transaction network. \\ 
& T6. Total value exchange (TVE) & The total exchanged value of a node, encapsulating its overall transfer value and economic activity. \\ 
& T7. Degree temporal density (DTD) & Distribution of transactions over time for each node, quantifying the frequency of transactions over time. \\ 
 & T8. Contract interaction ratio (CIR) & Proportion of transactions involving smart contract interactions relative to total transactions. \\ 
& T9. Peak neighbor transaction degree (PNTD) & Maximum transactional engagement a node has with any of its neighbors, reflecting the primary transactional relationship. \\  \hline
\multirow{9}{*}{\rotatebox{90}{Network structure (N)} }& N1. Node connectivity degree (NCD) & Measures how many different accounts a node has transacted with, reflecting its interaction level and centrality. \\ 
& N2. Indegree centrality (IC) & Node's importance as a recipient based on incoming transactions. \\ 
& N3. Outdegree centrality (OC) & Node's importance as a sender based on outgoing transactions. \\ 
 & N4. Degree centrality (DC) & Overall importance of a node based on the total number of transactions relative to other nodes. \\ 
& N5. Average neighbor degree centrality (ANDC) & Average degree centrality for a node's neighbors, reflecting the importance of surrounding nodes. \\ 
 & N6. Pagerank (PR) & Node's importance based on incoming links and the rank of linking nodes. \\ 
& N7. Square clustering coefficient (SCC) & Prevalence of square cycles (length four) in the graph, indicating complex transaction patterns. \\ 
 & N8. Maximal network reach (MNR) & Maximum distance from node \( v \) to any other node, reflecting relative remoteness or accessibility. \\ 
 & N9. Unique transactional pathways (UTP) & The number of unique paths from a node that reach its non-first-order neighbors without revisiting any node. \\ \hline
\end{tabular}}
\label{tab:nodefeatures}
\end{table*}

These features capture typical phishing patterns. A high AID reflects numerous incoming transactions typical of fraudulent fund collection, while a low AOD suggests limited outward transfers consistent with accumulation behavior. CIR and PNTD characterize the nature of transactions and relationships, which can also signal phishing. Phishing accounts often interact with inactive accounts, have few unique transaction paths, and rarely form closed transaction quadruplets, leading to low ANDC, UTP, and SCC. They also tend to show abnormally high PR values, whereas low scores may indicate peripheral involvement.
In addition to extracted features, we assign distinct numerical identifiers to heterogeneous addresses such as CA and EOA nodes, enhancing \graphname's node attributes and aiding phishing pattern learning.

\textbf{Edge features.}
In addition to node features, edge features play a critical role in capturing the interactions between accounts in \graphname. These features include transaction value, token type, transaction type, gas fees, and timestamp, each providing essential details for understanding the nature of the transactions. For each transaction, the transaction value indicates the amount of funds transferred, while the token type and transaction type help differentiate between different asset movements, such as Ether transfers, FT transfers, and NFT transfers. Gas fees represent the computational cost for executing the transaction, which can reveal atypical transaction behaviors, often associated with phishing activities. The timestamp adds temporal context, allowing the model to recognize transaction patterns and detect suspicious timing behaviors, such as rapid fund movements.

	\subsection{Temporal Graph Contrastive Learning}
	\label{sec:tgl}

Traditional GNNs, e.g., GCN~\cite{kipf2016semi}, GAT~\cite{brody2021attentive}, MPNN~\cite{gilmer2017neural}, and GraphSAGE~\cite{hamilton2017inductive},
	learn node and edge representations in simple homogeneous graphs by integrating neighbor features and structural information, which works for tasks like classification and link prediction.
To our knowledge, few studies have specifically explored node and edge representation learning on heterogeneous graphs.
To enable temporal graph learning in \graphname, we design PhishTGL, a self-supervised model that integrates temporal encoding, historical and neighbor aggregation, edge representation learning, and node-level contrastive learning.

	\emph{Temporal encoding.}
	Drawing inspiration from Monte Carlo integral~\cite{rahimi2007random} and TGAT~\cite{xu2020inductive}, our approach employs functional time encoding to map the time domain into a continuous, differentiable functional domain. This ensures compatibility with the self-attention mechanism and facilitates optimization via backpropagation. The functional temporal encoding is formulated as:
	\begin{align}
		\label{eq:tem_encoding}
		\nonumber \resizebox{.92\linewidth}{!}{	$t \mapsto \Phi_d(t):=\sqrt{\frac{1}{d}}\left[\cos \left(\omega_1 t\right), \sin \left(\omega_1 t\right), \ldots, \cos \left(\omega_d t\right), \sin \left(\omega_d t\right)\right],$}
	\end{align}
	where the time is projected into a $2d$-dimensional representation vector using the learnable weight $\omega_1,..\omega_n$.

	\emph{Aggregating historical node representation.}
	At any given time \( t \), each node \( v \) maintains a memory state \( s_v(t) \) characterizing the node's historical interactions.
	The memory is dynamically updated following each transaction and the involving nodes.
	For a specific transaction \( e(v,u,t_e) \) between nodes \( v \) and \( u \) at timestamp $t_e$, two messages are computed as follows:
	\begin{align}
		\nonumber	\mathbf{m}_{v}(t_e) & = \operatorname{msg}_{\mathrm{s}}\left(\mathbf{s}_{v}\left(t_e^{-}\right), \mathbf{s}_{u}\left(t_e^{-}\right), \Phi_d(t)(\Delta t), e(v,u,t_e) \right) , \\
		\nonumber	\mathbf{m}_{u}(t_e) & = \operatorname{msg}_{\mathrm{d}}\left(\mathbf{s}_{u}\left(t_e^{-}\right), \mathbf{s}_{v}\left(t_e^{-}\right), \Phi_d(t)(\Delta t), e(v,u,t_e) \right),
	\end{align}
	where \( s_{v}(t_e^-) \) denotes the memory of node \( v \) just before time \( t \), and \( \operatorname{msg}_s \) and \( \operatorname{msg}_d \) are learnable message functions.
$\Delta t$ is the time since the last interaction.
	Messages for the same node in a batch are aggregated to form a single message \( \bar{\mathbf{m}}_{v}(t) \) by taking the mean of the messages:
	\begin{equation}
		\bar{\mathbf{m}}_{v}(t_e) = \text{Agg}(\mathbf{m}_{v}(t_1), \ldots, \mathbf{m}_{v}(t_n)),
	\end{equation}
	where $t_1...t_n \leq t_e$.
	A gated recurrent unit (GRU) is employed to update the memory of the node, integrating new information from recent interactions. The updated memory state is given as Eq.~\ref{eq:memory}, reflecting the latest temporal dynamics.
	\begin{equation}
		\label{eq:memory}
		\mathbf{s}_{v} (t_e) = \text{GRU}(\bar{\mathbf{m}}_{v}(t_e), \mathbf{s}_{v}(t_e^-)), v \in \{ v,u\}.
	\end{equation}

	\emph{Aggregating neighbor node representation.}
	We employ a temporal attention-based strategy to aggregate the neighborhood temporal representations.
	The representation of node \( v \) at time \( t \) and \( l_{th}\) layer, \( \mathbf{h}^l_{v}(t) \), is formulated as:
	\begin{align}
		\mathbf{h}^l_{v}(t) = \operatorname{MultiHeadAtten}^l(\mathbf{q_v}^l(t),\mathbf{K_v}^l(t),\mathbf{V_v}^l(t)).
	\end{align}
	The output $\mathbf{h}^l_{v}(t)$ is a multi-head feature representation.
$\mathbf{q}^l(t)$, $\mathbf{K}^l(t)$, and $\mathbf{V}^l(t)$ are formulated as:
	\begin{align}
		\nonumber		\mathbf{q}_v^l(t) =  [\mathbf{h}_0^{(l-1)}(t) \|\Phi_{d}(0) ] \mathbf{W}_Q,
	\end{align}
	\begin{align}
		\nonumber	\resizebox{.92\linewidth}{!}{
		$\mathbf{K}_v^l(t) =  [\mathbf{z}_1^{(l-1)}(t) \|\mathbf{z}^l_{e_1}{(t_1)} \|\Phi_{d}(t-t_1) \dots  \mathbf{z}_i^{(l-1)}(t) \|\mathbf{z}^l_{e_k}{(t_k)} \|\Phi_{d}(t-t_i) \dots] \mathbf{W}_K$,}
	\end{align}
	\begin{align}
		\nonumber	\resizebox{.92\linewidth}{!}{
		$\mathbf{V}_v^l(t) =  [\mathbf{z}_1^{(l-1)}(t) \|\mathbf{z}^l_{e_1}{(t_1)} \|\Phi_{d}(t-t_1) \dots  \mathbf{z}_i^{(l-1)}(t) \|\mathbf{z}^l_{e_k}{(t_k)} \|\Phi_{d}(t-t_i) \dots] \mathbf{W}_V$,}
	\end{align}
	where $\mathbf{W}_Q$, $\mathbf{W}_K$, and $\mathbf{W}_V$ are learnable weights to capture the interactions between time encoding and node features.
$\|$ is the concatenation operation.
$\mathbf{z}^l_{e_k}$ is the representation of edge $e_k$, formulated in {TGL-IV}.
	The representation $\mathbf{z}_i^{(l-1)}(t)$ denotes the final representation of the $i_{th}$ neighbor of node $v$ at the previous layer.
	To capture complex feature interactions, we employ a feed-forward neural network (FFN) as per GraphSAGE, with $\mathbf{z}_v^{l}(t)=\operatorname{FFN}^{(l)}\left(\mathbf{z}_v^{(l-1)}(t) \|\mathbf{h}_v^{(l)}(t)\right)$.
	PhishTGL aggregates representation across L layers, accumulating information within L-hop temporal neighborhoods to construct the node's embedding in the TGL.

	\emph{Edge representation learning.}
	\sys is also designed to learn edge representation as follows.
	For a specific transaction edge \( e(v, u, t_e) \), PhishTGL \emph{first} aggregates representations of incoming neighboring edges with the same recipient \( u \) and timestamps earlier than \( t_e \), formulated as:
	\begin{align}
		\resizebox{.89\linewidth}{!}{$
				h_{IN(e(v, u, t_e))}^l = \operatorname{Agg}\left(\{z_{(v', u, t_{e'})}^l : \forall u' \in \mathcal{V}, e(v', u, t_{e'}) \in \mathcal{E}, t_{e'} < t_e\}\right)$}
	\end{align}
	The intuition is that the concentration of incoming transactions to a single account is indicative of malicious activities, particularly phishing transactions.
	\emph{Second}, PhishTGL incorporates the representations of two nodes, given as Eq.~\ref{eq:edgeuv}.
	The intuition is that incorporating sender and recipient node representations enhances understanding of transaction context to better understand phishing activities.
	\begin{align}
		\label{eq:edgeuv}
		h_{uv(t_e)}^l = \operatorname{Agg}(z_u^l(t_e),z_v^l(t_e) )
	\end{align}
	\emph{Third}, PhishTGL aggregates edge representations and propagates them through multiple graph layers, thereby enriching node and edge representations with deeper structural insights.
	It is formulated as:
	\begin{align}
		\label{eq:edgef}
		z_{e(v, u, t_e)}^l = \operatorname{Agg}( z_{e(v, u, t_e)}^{l-1} \| h_{uv(t_e)} \| h_{IN(e(v, u, t_e))}^l ).
	\end{align}

	\emph{Node-level contrastive learning.}
	PhishTGL model is trained in a self-supervised strategy by applying node-level graph contrastive learning.
 The key idea is to maximize node representation consistency across graph views. We first generate two subgraphs \( g_1 \) and \( g_2 \) by perturbing the original transaction graph.
	Then we use the PhishTGL to derive node representations for these subgraphs, and next perform contrastive learning by comparing node consistency between views.
	To create distinct views for comparison, we apply structural and feature perturbing of the original graph to generate subgraphs:
	\begin{inditemize}
		\item Edges are randomly removed using a Bernoulli mask $ \mathds{1}_e\in \{ 0,1\}^{|\mathcal{E}|} $, with the perturbed edges as \( \tilde{\mathcal{E}} = \mathcal{E} \circ \mathds{1}_e \), where $\circ$ is element-wide product.
		\item Node features are masked with zeros randomly, i.e., $ \tilde{\mathcal{X}_v} = \mathcal{X}_v \circ \mathds{1}_v$, where Bernoulli mask $\mathds{1}_v \in \{ 0,1\}^{|\mathcal{V}|}$.
	\end{inditemize}

	For node \( i\), the representation at time $t$ in one view $g_1$ serves as the anchor, e.g., \( \mathbf{z}^{g_1}_{i}(t) \).
	The representation of the node $v_i$ at the same time in the other view $g_2$ is regarded as the positive sample, denoted as \( \mathbf{z}^{g_2}_{i}(t)\).
	The representations of all other nodes act as negative samples.
	Given a positive pair, we naturally define negative samples as all other nodes in the two views.
	Thus, the pairwise objective for each positive pair $(\mathbf{z}^{g_1}_{i}(t), \mathbf{z}^{g_2}_{i}(t))$ can be formulated as:
	\begin{equation}
		\resizebox{.9\linewidth}{!}{$
			\ell_t(\mathbf{z}^{g_1}_{i}, \mathbf{z}^{g_2}_{i}) = \log \frac{e^{\theta_t(\mathbf{z}^{g_1}_{i}, \mathbf{z}^{g_2}_{i}) / \tau}}
			{e^{\theta_t(\mathbf{z}^{g_1}_{i}, \mathbf{z}^{g_2}_{i}) / \tau}
			+ \sum_{k} \mathds{1}_{[k\neq i]}e^{\theta_t(\mathbf{z}^{g_1}_{i}, \mathbf{z}^{g_2}_{k}) / \tau}
			+ \sum_{k} \mathds{1}_{[k\neq i]}e^{\theta_t(\mathbf{z}^{g_1}_{i}, \mathbf{z}^{g_1}_{k}) / \tau}}$}
	\end{equation}
	where \(\theta_t(\mathbf{z}^{g_1}_{i}, \mathbf{z}^{g_2}_{i})  =s(p(\mathbf{z}^{g_1}_{i}(t)),p(\mathbf{z}^{g_2}_{i}(t))) \),
	is the projected cosine similarity between representation of $\mathbf{z}^{g_1}_{i}(t)$
	and $\mathbf{z}^{g_2}_{i}(t)$  at timestamp $t$.
	Projector $p(\cdot )$ is implemented with a two-layer FNN.
$\mathds{1}_{[k\neq i]}\in \{0,1\}$ is an indicator function that equals to 1 if $k\neq i$.
	\( \tau \) is the temperature parameter, and the overall objective is the average of all positive pairs.
$e^{\theta_t(a_i, v_i) / \tau}$ is the positive pair.
$\sum_{k} \mathds{1}_{[k\neq i]}e^{\theta_t(\mathbf{z}^{g_1}_{i}, \mathbf{z}^{g_2}_{k}) / \tau} $ is the inter-view negative pair, and
$\sum_{k} \mathds{1}_{[k\neq i]}e^{\theta_t(\mathbf{z}^{g_1}_{i}, \mathbf{z}^{g_1}_{k}) / \tau} $ is the intra-view negative pair.
	Since two views are symmetric, the loss for another view $g_1$ is similar as $(\mathbf{z}^{g_1}_{i}(t), \mathbf{z}^{g_2}_{i}(t))$.
	The overall objective is defined as average overall positive pairs (Eq.~\ref{eq:loss}), which is trained to be maximized using stochastic gradient ascent (SGA).
	\begin{align}
		\label{eq:loss}
		\mathcal{L} = \frac{1}{2N} \sum_{i=1}^{N} [\ell_t((\mathbf{z}^{g_1}_{i}, \mathbf{z}^{g_2}_{i})) + \ell_t((\mathbf{z}^{g_2}_{i}, \mathbf{z}^{g_1}_{i}))]
	\end{align}
	Finally, the trained PhishTGL model is used for generating the node and edge representations.

	\subsection{Phishing Detection}

Using the PhishTGL model pre-trained in a transductive setting, \sys predicts whether new nodes and edges in \graphname are phishing or non-phishing based on their learned representations. These representations are mapped to two classes, and the phishing detector assigns the label according to the higher logit value. For classification, we employ LightGBM~\cite{LightGBM}, a gradient boosting decision tree framework that efficiently trains ensembles of weak classifiers to optimize detection performance.

\section{Evaluation}

In this section, we present the evaluation of \sys on a compiled dataset, its real-world deployment performance, and case studies.
	
	\subsection{Evaluation on a Compiled Dataset}

To evaluate phishing detection on Ethereum, we compiled a high-confidence ground-truth dataset by collecting verified phishing and non-phishing addresses and then sampling their associated transaction histories. 

We constructed a high-confidence phishing set by randomly sampling 1,000 \emph{verified phishing addresses} from two leading Web3 security service providers that continuously monitor on-chain scams and maintain operational threat intelligence databases. These providers collect phishing addresses from real-world incident reports, victim complaints, phishing website investigations, and internal on-chain tracing. We selected them because they offer broad Ethereum coverage, maintain independently curated labels, and are widely used in practice, which helps reduce source-specific bias. The sampled addresses span incidents from \textit{[start year]} to \textit{[end year]}. To further improve label reliability, we asked ScamSniffer's~\cite{scamsniffer} risk auditors to re-verify the sampled addresses through: (i) direct victim reports, (ii) cross-validation with multiple independent platforms, including BlockSec~\cite{blocksec}, CertiK~\cite{certik}, GoPlus~\cite{goplus}, SlowMist~\cite{slowmist}, and Etherscan Labelcloud~\cite{Label_Word}, and (iii) manual review by experienced auditors.

We used 1,000 phishing addresses because our goal was to build a \emph{high-confidence} positive set rather than a larger but noisier one. In practice, verified phishing labels are much harder to obtain than benign labels, and increasing the positive set would introduce more uncertain or weakly supported cases. For the benign class, we randomly sampled 160,658 recently active addresses that were not associated with any known phishing reports and further checked them against the same platforms. This process yields a reliable dataset while maintaining diversity and minimizing labeling noise.

\begin{figure}[!t]
		\centering
		\subfloat[]{\includegraphics[width = 0.32\linewidth]{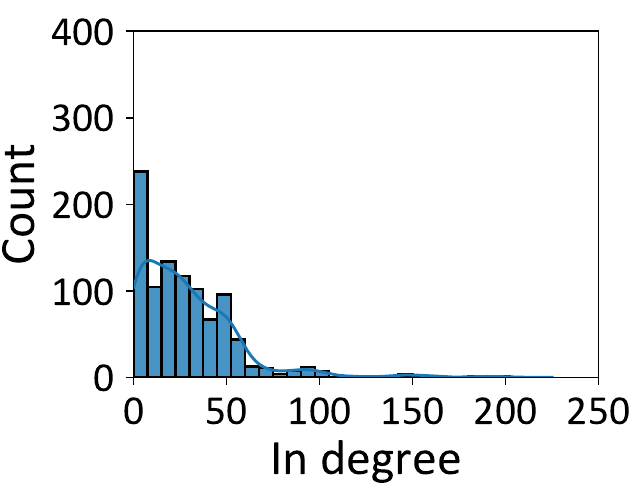}} 
		\subfloat[]{\includegraphics[width = 0.32\linewidth]{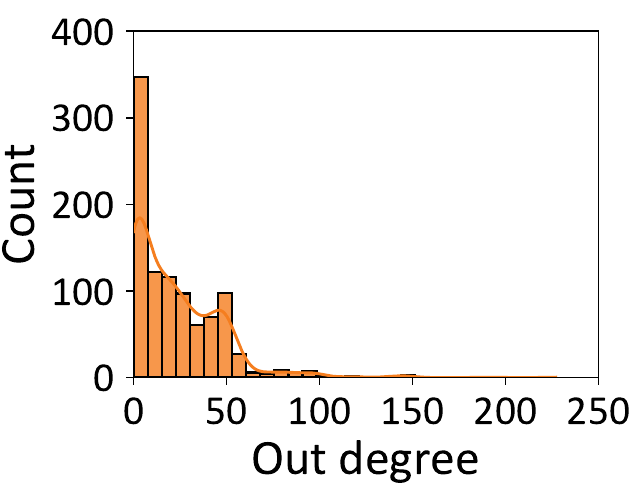}}
		\subfloat[]{\includegraphics[width = 0.32\linewidth]{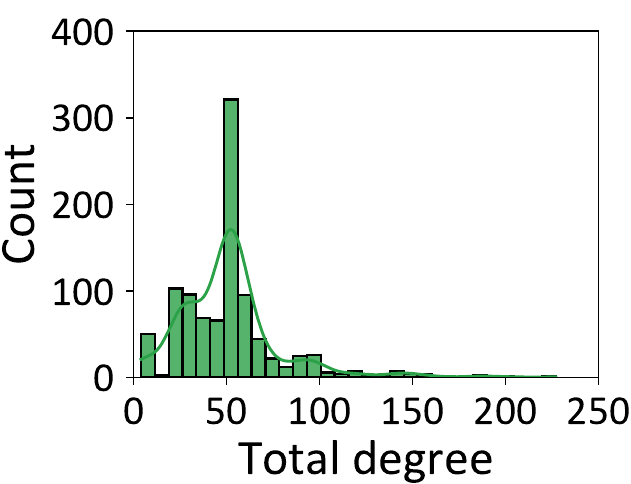}}
		\caption{Distribution of phishing addresses' indegree (a), outdegree (b), and total degree (c)}
		\label{fig:distribution_deg} 	
	\end{figure}

	We then sampled transaction dataset of these phishing addresses and non-phishing addresses between 2015 and March 2023. 
	In detail, we extended first- and second-order neighbors of their transactions as our dataset.
	We limited the dataset to include up to 50 transactions per address.
	Based on these addresses,  we finally extracted the transaction dataset comprising a total of {416,541} transactions, including {48,135} phishing transactions and {368,406} non-phishing transactions.
	For each transaction, we extracted sender and recipient addresses, timestamps, values, gas fees, transaction types, and token names, respectively.
	This dataset serves as a reliable ground-truth resource for Ethereum phishing detection, providing a fair, generalizable, and reproducible benchmark for evaluating detection methods.
	Figure~\ref{fig:distribution_deg} shows the distribution of phishing addresses' in degree, out degree, and total degree, respectively.
	The average in degree, out degree, and total degree of the phishing addresses were 22.63±29.16, 22.63±24.43, and 51.75±29.42, respectively.

	\subsubsection{Evaluation Setup}

	\textbf{Hyperparameters.}
	For the PhishTGL model, the number of graph layers is 2.
	The learning rate and the batch size are 0.0001 and 256 in training PhishTGL.
	Temperature $\tau$ for loss function $\mathcal{L}$ is 1.
	Two attention layers are used with the number of attention heads as 8.
	The dimension of node memory and temporal representation is 128.
	We set the probability $p$ of the two Bernoulli distributions in $\mathds{1}_v$ and $\mathds{1}_e$ as 20\%, i.e., randomly removing 20\% edges and masking 20\% node features.
	We employed a mean aggregator for all $\operatorname{Agg}$ operators, i.e., calculating element-wise means of the vectors to be aggregated.
	Parameter $d$ of node feature PR is 0.85.
	We varied the representation dimension of node and edge in \{16, 32, 64, \underline{128}\}, where \underline{\quad} denotes the default evaluation setting.
	We also varied neighbors in \{5, 10, 15, \underline{20}\} for graph learning.
	The hyperparameters of lightGBM for downstream node and edge prediction were decided via cross-validation.
	We set $num\_leave$  in node prediction and edge prediction as 127 and 63, respectively. The $feature\_fraction$ was set to 0.9 to speed up training and prevent over-fitting.
The Learning rate for lightGBM was 0.08 and the train-test split was 7:3.
To train a robust detection model on such an imbalanced dataset, we performed random edge upsampling and random node downsampling with a sampling ratio of 0.8.
For each setup, we performed the 10-fold cross-validation and evaluated average performance.
All evaluations were conducted on the device equipped with an Intel(R) Xeon(R) Gold 6133 CPU @ 2.50GHz and an NVIDIA GeForce RTX 4090 GPU.

\textbf{Graph models for comparison.}
We compared the performance of \sys with the following three graph learning models:
(i) \emph{DeepWalk}~\cite{perozzi2014deepwalk} utilizes random walks on a graph to generate node sequences, treating them like sentences in natural language.
These generated node sequences are then processed through a skip-gram model~\cite{mikolov2013distributed} to learn node representations that reflect the graph's neighborhood structure;
(ii) \emph{GraphSAGE}~\cite{hamilton2017inductive} is a static GNN-based approach.
It employs node feature aggregation from a node's local neighborhood for embedding, and uses convolutional neural networks to encode the aggregated information to capture the context of each node in its network locality; and
(iii) \emph{TGAT}~\cite{xu2020inductive} is state-of-the-art (SOTA) dynamic GNN model inspired by graph attention model, which integrates a time-aware attention mechanism with neighbor embedding aggregating, to represent the evolution of nodes and their relationships.

While the three methods are previously designed for node embedding learning, we made some modifications for edge embedding learning in our task.
For DeepWalk and GraphSAGE, we computed edge embeddings by concatenating the representation of the connected nodes and the edge features.
The edge representation in TGAT, was adapted to concatenate the temporal representation of the two connected nodes, edge features, and its temporal representation.
In training comparison graph models, we employed the same parameters as in our method for the three models.

\textbf{Metrics.}
We used the following metrics in our evaluation.
Precision measures the proportion of true positives (phishing) among all positive predictions.
Recall measures the proportion of true positives among all real positives.
FNR and FPR measure the ratios of incorrect detections, i.e., FNR for missed positives, and FPR for false alarms.
F1-score, harmonic mean of precision and recall, provides a balanced view of the model's overall accuracy, especially useful for unbalanced class distributions.
AUC is used to estimate the probability that prediction scores of phishing cases are higher than the non-phishing ones.
BAC, the mean of the true positive rate and true negative rate, is a widely used metric for evaluating on imbalanced dataset.

\begin{table}[!t]
	\centering
	\small
	\caption{Mean/standard deviation of metric in detecting phishing transactions and accounts}	
	\begin{tabular}{lll}
		\hline
		{Metric}    & {Phishing transaction (\%)} & {Phishing account (\%)} \\
		\hline
		{Precision} & 99.25/0.02                        & 92.24/0.78                    \\
		{Recall}    & 77.81/0.65                        & 96.24/0.94                    \\
		{FPR}       & 0.59/0.02                         & 8.10/0.88                     \\
		{FNR}       & 22.19/ 0.65                       & 3.76/0.94                     \\
		{F1}        & 87.23/0.40                        & 94.19/0.49                    \\
		{AUC}       & 98.43/0.04                        & 98.03/0.38                    \\
		{BAC}       & {88.61/0.32 }              & {94.07/0.60}           \\
		\hline
	\end{tabular}
	\label{tab:performance_overall}
\end{table}
\subsubsection{Overall Performance}

\begin{figure*}[!t]
	\centering
	\subfloat[Detecting phishing transactions]{\includegraphics[width = 0.28\linewidth]{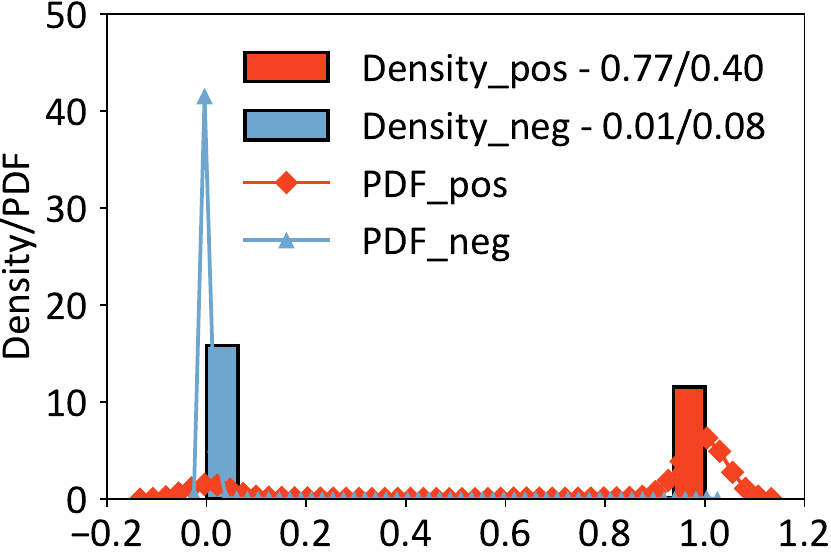}} \hspace{2mm}
	\subfloat[Detecting phishing accounts]{\includegraphics[width = 0.28\linewidth]{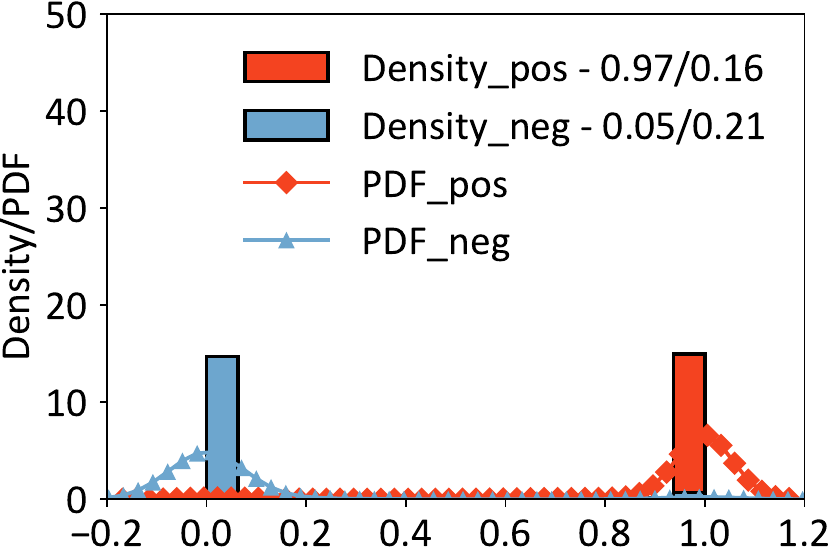}}\hspace{2mm}
	\subfloat[W/o node features ]{\includegraphics[width = 0.28\linewidth]{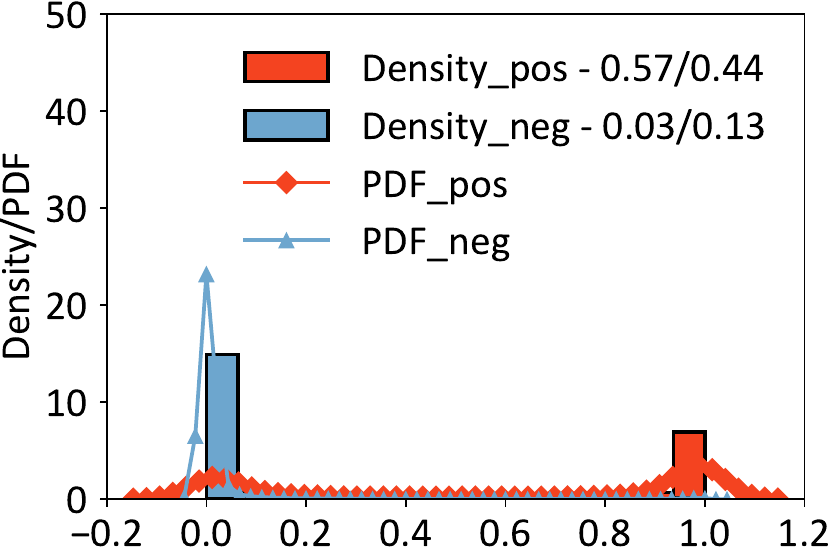}}	
	\caption{PDF and density of prediction scores in detecting phishing transactions (a) and accounts (b), and detecting phishing transactions under the model  without extracted node features (c)}
	\label{fig:edge_detect}	
\end{figure*}

We evaluated \sys's performance in two tasks, e.g., detection of phishing accounts and transactions.
We set the dimension of node representation as 128, and the number of sampled neighbors is 20.
Table~\ref{tab:performance_overall} reports the overall mean precision, recall, F1 score, and AUC in detecting phishing accounts and transactions.
For detecting phishing transactions, \sys achieves a robust precision of 99.25\% (±0.02),  recall of  77.81\% (±0.65), FPR of 0.59\%,
FNR of 22.19\%, signifying an adept ability to correctly identify phishing accounts. The F1 score of 87.23\% (±0.40) and an AUC of 98.43\% (±0.04) indicate a high level of overall accuracy.
Phishing accounts detection shows a slightly lower precision of 92.24\% (±0.78) but an impressive recall of 96.24\% (±0.94), FPR of 8.10\%,
FNR of 3.76\%, reflecting the system's strength in capturing the majority of phishing transactions.
F1 score is 94.19\% (±0.49), with an AUC of 98.03\% (±0.38), showcasing \sys's effectiveness in identifying phishing transactions, albeit with a slightly larger margin of error.
The results demonstrate \sys's proficient and reliable performance in detecting phishing activities.
Figure~\ref{fig:edge_detect}(a) and (b) present the probability density function (PDF) and density of prediction score in detecting phishing transactions and accounts.

\subsubsection{Impact of Representation Dimensions}

\begin{figure}[!t]
	\centering
	\small
	\includegraphics[width = 0.8\linewidth]{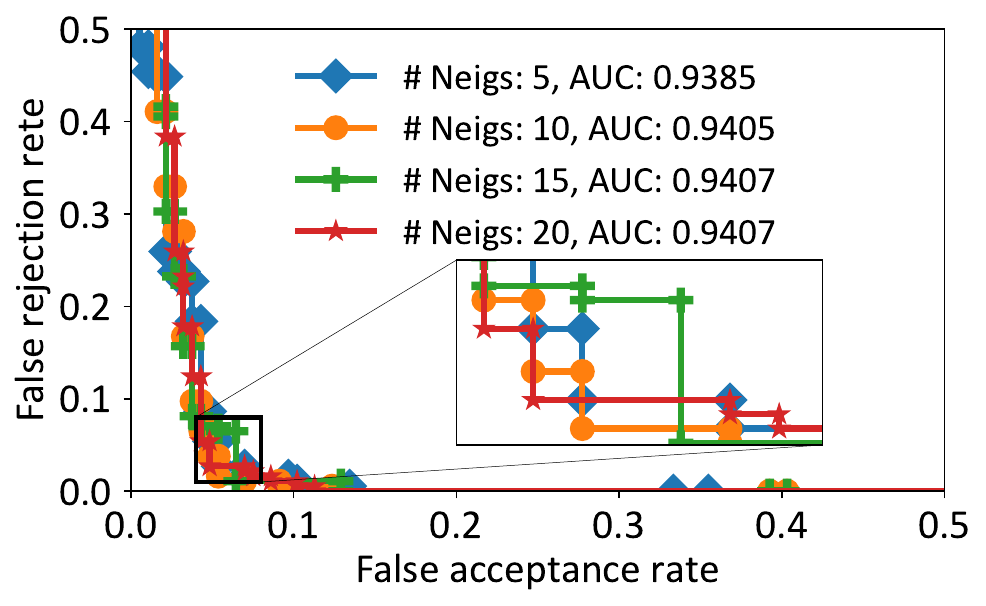}	
	\caption{ROC curves under different sampled neighbors}
	\label{fig:roc}
\end{figure}

Table~\ref{tab:performance_edge_pre} and Table~\ref{tab:performance_node_pre} report performance under different representation dimensionality on detecting phishing transactions and accounts. The results show that a 128-dimensional representation delivers the best performance for both tasks, balancing precision, recall, and computational efficiency. For phishing transactions, precision reaches 99.25\%, recall 77.81\%, F1 score 87.23\%, and AUC 98.43\%, as shown in the receiver operating characteristic (ROC) curves in Figure~\ref{fig:roc}. For phishing accounts, the 128-dimensional representation achieves a precision of 92.24\%, recall of 96.24\%, F1 score of 94.19\%, and AUC of 98.03\%. These results indicate that while increasing the dimensionality enhances representation richness, overly high dimensions may not lead to significant performance gains and could introduce computational overhead.

\begin{table}[!t]
	\centering
	\small
	\caption{Mean/standard deviation of precision, recall, F1, AUC under different representation dimensions and sampled neighbors in detecting phishing transactions} 	
	\begin{tabular}{p{22pt}p{15pt}cccccccr}
		\hline
		{\#Dim}       & {\#Neig} & {Precision (\%)} & {Recall (\%)} & {F1 (\%)}      & {AUC (\%)}    \\
		\hline
				\multirow{4}{*}{16}  & 5               & 98.56/0.13             & 70.09/0.42          & 81.92/0.28           & 97.35/0.11          \\
		                     & 10              & 98.51/0.09             & 70.52/0.35          & 82.20/0.22           & 97.35/0.07          \\
		                     & 15              & 98.24/0.04             & 69.94/0.75          & 81.71/0.51           & 97.39/0.04          \\
		                     & 20              & 98.72/0.14             & 69.37/0.77          & 81.48/0.51           & 97.31/0.10          \\
							 \hline
							 \multirow{4}{*}{32}  & 5               & 98.62/0.10             & 72.15/0.61          & 83.33/0.41           & 98.15/0.03          \\
		                     & 10              & 98.60/0.03             & 71.56/0.11          & 82.93/0.08           & 98.21/0.03          \\
		                     & 15              & 98.72/0.07             & 71.09/0.55          & 82.66/0.38           & 98.30/0.03          \\
		                     & 20              & 98.52/0.05             & 68.96/0.10          & 81.13/0.08           & 96.97/0.07          \\
							 \hline
							 \multirow{4}{*}{64}  & 5               & 99.14/0.06             & 76.23/0.18          & 86.19/0.13           & 97.97/0.07          \\
		                     & 10              & 99.15/0.01             & 76.83/0.50          & 86.58/0.33           & 97.95/0.01          \\
		                     & 15              & 99.16/0.03             & 76.75/0.65          & 86.53/0.42           & 98.02/0.05          \\
		                     & 20              & 99.13/0.03             & 75.55/0.40          & 85.75/0.26           & 98.08/0.05          \\
							 \hline
							 \multirow{4}{*}{128} & 5               & 99.11/0.07             & 77.02/0.24          & 86.68/0.16           & 98.10/0.01          \\
		                     & 10              & 99.19/0.07             & 78.51/0.42          & 87.65/0.26           & 98.33/0.02          \\
		                     & 15              & 98.98/0.06             & 77.52/0.23          & 86.94/0.15           & 98.15/0.01          \\
		                     & 20              & {99.25/0.02}    & {77.81/0.65} & {87.23/0.40	} & {98.43/0.04} \\
		\hline
	\end{tabular}
	\label{tab:performance_edge_pre}	
\end{table}

\begin{table}[!t]
	\centering
	\small
	\caption{Performance under different representation dimensions and sampled neighbors in detecting accounts}	
	\begin{tabular}{p{22pt}p{15pt}cccccccr}
		\hline
		{\#Dim}                & {\#Neig} & {Precision (\%)} & {Recall (\%)}   & {F1 (\%)}     & {AUC (\%)}    \\ 			
			\hline
		\multirow{4}{*}{16}           & 5               & 89.76/1.68             & 95.46/0.54            & 92.51/0.86          & 92.28/0.84          \\
		                              & 10              & 90.58/2.28             & 95.16/1.15            & 92.80/1.60          & 92.61/1.74          \\
		                              & 15              & 89.76/1.12             & 94.50/0.46            & 92.06/0.51          & 91.84/0.67          \\
		                              & 20              & 90.05/0.99             & 94.71/1.28            & 92.31/0.68          & 92.11/0.72          \\
		\hline
		\multirow{4}{*}{32}           & 5               & 91.36/1.66             & 95.60/1.28            & 93.41/0.85          & 93.26/1.02          \\
		                              & 10              & 91.18/1.29             & 95.72/1.65            & 93.37/0.58          & 93.22/0.82          \\
		                              & 15              & 91.31/0.41             & 95.39/1.48            & 93.30/0.75          & 93.16/0.94          \\
		                              & 20              & 91.27/0.84             & 96.05/1.72            & 93.59/0.94          & 93.43/1.12          \\
									  \hline
		\multirow{4}{*}{64}           & 5               & 92.39/0.92             & 96.77/0.98            & 94.53/0.52          & 94.38/0.74          \\
		                              & 10              & 92.03/1.23             & 97.09/0.36            & 94.49/0.60          & 94.33/0.76          \\\
		                              & 15              & 91.80/1.65             & 96.25/1.07            & 93.96/0.80          & 93.80/1.06          \\
		                              & 20              & 91.13/0.95             & 96.35/1.06            & 93.66/0.61          & 93.46/0.89          \\
									  \hline
		{\multirow{4}{*}{128}} & 5               & 92.12/1.14             & 95.91/0.77            & 93.97/0.53          & 93.85/0.66          \\
		                              & 10              & 91.82/0.92             & 96.78/0.98            & 94.23/0.80          & 94.05/1.02          \\
		                              & 15              & 92.16/0.97             & 96.35/0.96            & 94.20/0.64          & 94.07/0.78          \\
		                              & {20}     & {92.24/0.78}    & {96.24/0.94  } & {94.19/0.49} & {98.03/0.38} \\
									  \hline
	\end{tabular}
	\label{tab:performance_node_pre}	
\end{table}

\subsubsection{Impact of Sampled Neighbors}

The number of sampled neighbors also influences the model's performance, but increasing the number of neighbors does not necessarily lead to improvements. Table~\ref{tab:performance_edge_pre} shows that while recall remains relatively high at 20 neighbors (77.52\%), there is no significant improvement compared to fewer neighbors, and precision slightly decreases. This suggests that including more neighbors may introduce redundant or irrelevant information, which can dilute the quality of feature aggregation. Similarly, Table~\ref{tab:performance_node_pre} highlights that for phishing account detection, increasing the number of neighbors slightly increases recall but consistently lowers precision, as seen with 20 neighbors achieving 96.24\% recall but reduced precision at 92.24\%.
Figure~\ref{fig:roc} illustrates minimal differences in ROC curves across different neighbor counts when the representation dimension is fixed at 128, indicating that higher neighbor counts do not meaningfully enhance the model's ability to distinguish phishing activities. This limitation may arise from distant or weakly connected neighbors introducing noise rather than useful context, resulting in diminishing returns and highlighting the importance of focusing on a relevant subset of neighbors.

\subsubsection{Effectiveness of Node Features in \graphname}

\begin{figure*}[!t]
	\centering
	\subfloat[]{\includegraphics[width = 0.24\linewidth]{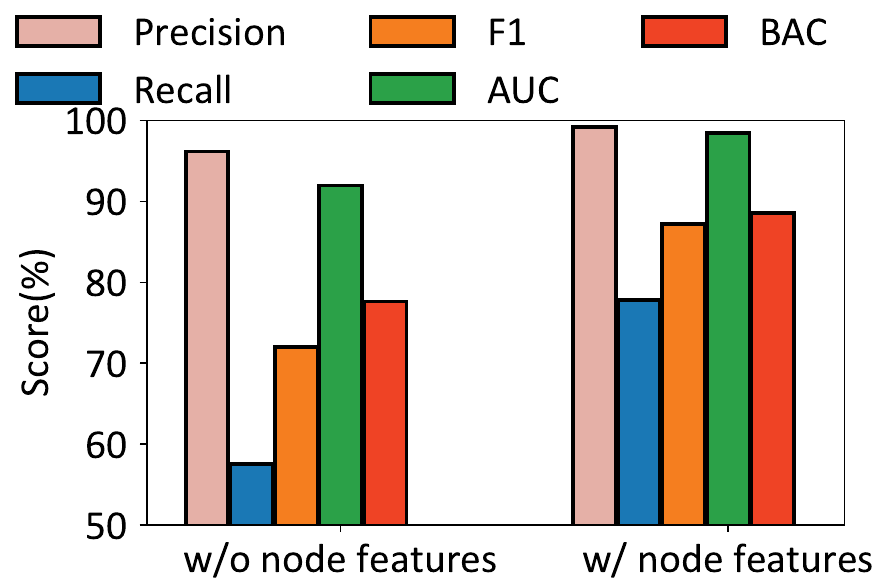}} \hspace{1mm}
	\subfloat[]{\includegraphics[width = 0.24\linewidth]{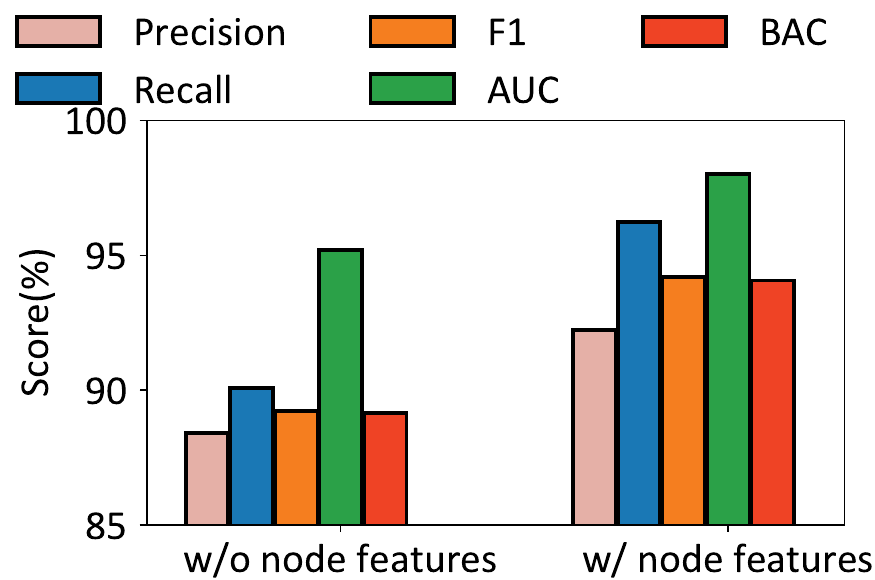}} \hspace{1mm}
	\subfloat[]{\includegraphics[width = 0.24\linewidth]{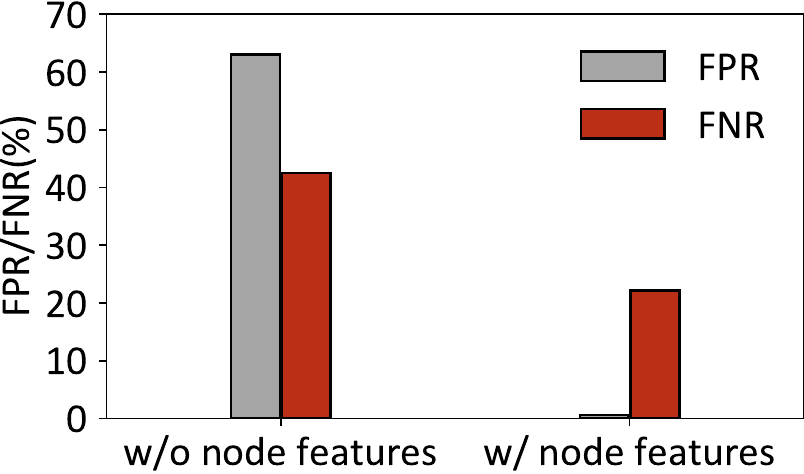}}\hspace{1mm}
	\subfloat[]{\includegraphics[width = 0.24\linewidth]{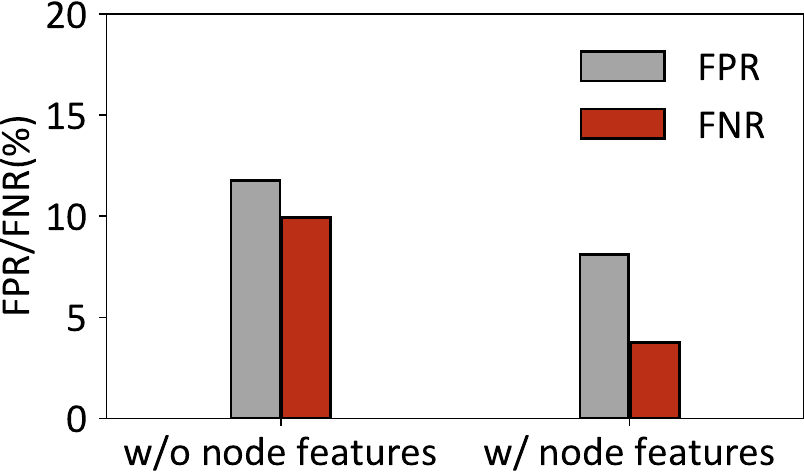}}
	\caption{Precision, recall, F1, AUC, and BAC in detecting phishing transactions (a) and accounts (b) under the model with or without extracted node features.
		FPR and FNR of detecting phishing transactions (c) and accounts (d)}
	\label{fig:performance_node_fea}
\end{figure*}

Figure~\ref{fig:performance_node_fea} shows that node features in \graphname significantly impact phishing transaction detection. Without node features, the model achieves FPR at 11.77\%, FNR at 9.93\%, F1 score at 89.22\%, and BAC at 89.15\% in detecting phishing accounts.
However, incorporating node features substantially improves the model's performance, evidenced by the enhanced metrics: FPR is improved to  8.10\%, FNR to 3.76\%, F1 score at 94.19\%, and BAC at 94.07\%.
While in the task of detecting phishing transactions, incorporating node features improves the FPR of 63\%, FNR of 42.46\%, F1 of 72.01\%, and BAC of 77.63\%, to the FPR of 0.59\%, FNR of 22.19\%, F1 of 87.23\%, and BAC of 88.61\%.
The reduced FPR and FNR highlight improved phishing detection accuracy, while low standard deviations demonstrate consistent and robust performance across runs.
In detail, Figure~\ref{fig:edge_detect} (c) presents prediction scores' PDF and density under the model without extracted features.

\subsubsection{Comparison with SOTA Graph Models}
Table~\ref{tab:baselines} presents performance under different graph learning models in detecting phishing transactions.
Results demonstrate the superior performance of \sys in phishing transaction detection.
DeepWalk shows the lowest performance with a precision of 51.11\%, recall of 41.62\%, and an F1 score of 45.88\%, indicating limited effectiveness in capturing phishing patterns. GraphSAGE, while better than DeepWalk, still shows limitations with a precision of 82.48\%, a low recall of 16.76\%, and an F1 score of 27.86\%. TGAT performs significantly better with precision at 98.32\%, recall at 70.93\%, and F1 at 82.41\%. However, \sys outperforms all with a precision of 99.25\%, recall at 77.81\%, F1 score at 87.23\%, and AUC at 98.43\%, highlighting its robustness in accurately identifying phishing accounts.

\begin{table}[!t]
	\centering
	\small
	\caption{Performance under different graph models in detecting phishing transactions}	
	\begin{tabular}{p{59pt}p{38pt}lll}
		\hline
		{Method}                                 & {Precision (\%)} & {Recall (\%)} & {F1 (\%)} & {AUC (\%)} \\
		\hline
		{DeepWalk}~\cite{perozzi2014deepwalk}    & 51.11                  & 41.62               & 45.88           & 50.90            \\
		{GraphSAGE}~\cite{hamilton2017inductive} & 82.48                  & 16.76               & 27.86           & 56.61            \\
		{TGAT}~\cite{xu2020inductive}            & 98.32                  & 70.93               & 82.41           & 84.86            \\\hline
		{\sys}                             & 99.25                  & 77.81               & 87.23           & 98.43            \\
		\hline
	\end{tabular}
	\label{tab:baselines}
\end{table}

Table~\ref{tab:baselines_accounts} provides the performance under different graph learning models for detecting phishing accounts. DeepWalk achieves a precision of 68.03\%, recall of 54.41\%, and an F1 score of 60.46\%, showing moderate performance. GraphSAGE improves slightly with a precision of 73.10\%, recall of 50.96\%, and an F1 score of 60.06\%. TGAT demonstrates better results with a precision of 78.78\%, recall of 65.34\%, and an F1 score of 71.43\%. \sys significantly outperforms these models with a precision of 92.24\%, recall of 96.24\%, an F1 score of 94.19\%, and an AUC of 98.03\%, underscoring its effectiveness in detecting phishing accounts.

\begin{table}[!t]
	\centering
	\small
	\caption{Performance under different graph learning models in detecting phishing accounts}
	\begin{tabular}{p{59pt}p{45pt}lll}
		\hline
		{Method}                                 & {Precision (\%)} & {Recall (\%)} & {F1 (\%)} & {AUC (\%)} \\
		\hline
		{DeepWalk}~\cite{perozzi2014deepwalk}    & 68.03                  & 54.41               & 60.46           & 70.03            \\
		{GraphSAGE}~\cite{hamilton2017inductive} & 73.10                  & 50.96               & 60.06           & 70.15            \\
		{TGAT}~\cite{xu2020inductive}            & 78.78                  & 65.34               & 71.43           & 77.66            \\\hline
		{\sys}                             & 92.24                  & 96.24               & 94.19           & 98.03            \\
		\hline
	\end{tabular}
	\label{tab:baselines_accounts}	
\end{table}

\subsubsection{Comparing with SOTA Detections}

We compare \sys with three representative and open-sourced phishing detection baselines, i.e., TTAGN~\cite{li2022ttagn}, TGC~\cite{li2023tgc}, and Trans2vec~\cite{wu2020phishers}. As shown in Table~\ref{tab:comparing_transactions}, \sys delivers the best overall performance for phishing transaction detection. Although its precision (99.25\%) is slightly below that of Trans2vec (99.68\%), \sys achieves a much stronger balance between precision and coverage, substantially improving recall to 77.81\%, compared with 71.28\% for TTAGN, 68.71\% for TGC, and only 39.45\% for Trans2vec. This advantage is further reflected in its lower false negative rate (22.19\%), which is markedly better than TTAGN (28.72\%), TGC (31.29\%), and especially Trans2vec (60.55\%). As a result, \sys obtains the highest F1 score (87.23\%), AUC (98.43\%), and BAC (88.61\%). These results indicate that \sys is not merely conservative in prediction, but is substantially more capable of identifying malicious transactions while maintaining a very low false positive rate (0.59\%).

A similar trend is observed for phishing account detection. Table~\ref{tab:comparing_accounts} shows that \sys consistently achieves the strongest and most balanced results across all methods, yielding the highest recall (96.24\%), the lowest false negative rate (3.76\%), the best F1 score (94.19\%), and the highest AUC (98.03\%). Compared with TTAGN and TGC, \sys improves both detection completeness and overall robustness, while also clearly outperforming Trans2vec by a large margin. Notably, \sys maintains competitive precision (92.24\%) and a relatively low false positive rate (8.10\%), which remain better than TTAGN (11.15\%) and Trans2vec (18.11\%). Overall, these results suggest that \sys is more effective at capturing phishing accounts with diverse and evolving transaction behaviors, and that its temporal and heterogeneous modeling provides a clear advantage over existing static and semi-dynamic baselines.

\subsubsection{Generalizing to Unseen Nodes}
\label{app:generalizing_unseen}

To further evaluate the generalization capability of \sys on emerging phishing activities, we consider a stricter unseen-node setting where the training and testing sets contain no overlapping nodes. We chronologically split the dataset, using earlier transactions for training and later transactions for testing, so as to better reflect the real-world deployment scenario in which new accounts continuously appear over time. This setting is substantially more challenging than standard evaluation, because the model cannot rely on previously observed node identities or historical embeddings of test nodes. Instead, it must infer phishing behaviors from temporal interaction patterns, structural contexts, and heterogeneous transaction semantics. Therefore, this experiment provides a more realistic and rigorous assessment of whether \sys can detect phishing activities in evolving Ethereum transaction graphs.

\begin{figure}[!t]
	\centering
	\subfloat[]{\includegraphics[width = 0.48\linewidth]{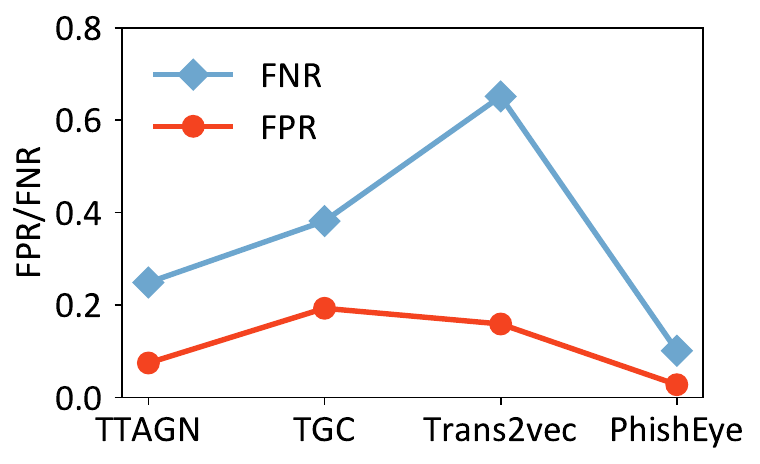}} \hfil
	\subfloat[]{\includegraphics[width = 0.48\linewidth]{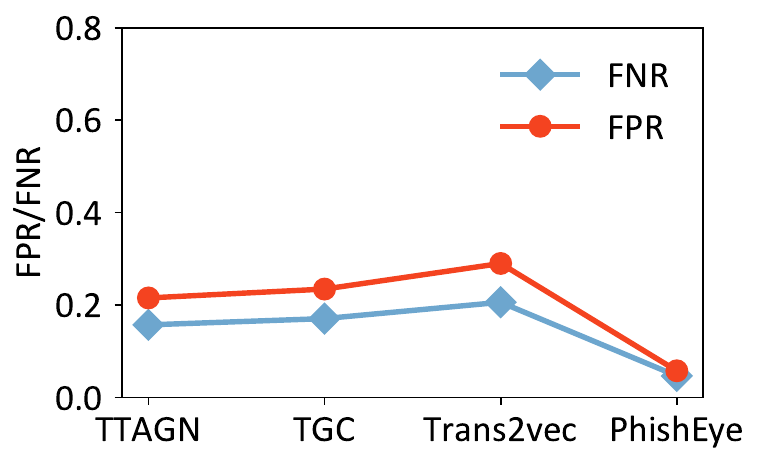}}
	\caption{FNR/FPR of different methods in detecting phishing transactions (a) and accounts (b) under an unseen-node setting}
	\label{fig:useen_error}
\end{figure}

As shown in Figure~\ref{fig:useen_error}, \sys consistently achieves the lowest error rates for both phishing transaction detection and phishing account detection under this setting. For transaction detection, \sys reduces the error to 10.09\% FNR and 2.74\% FPR, substantially outperforming TTAGN (24.86\%/7.44\%), TGC (38.16\%/19.28\%), and Trans2vec (65.13\%/15.88\%). A similar trend holds for account detection, where \sys achieves only 4.64\% FNR and 5.75\% FPR, compared with 15.69\%/21.52\% for TTAGN, 17.05\%/23.44\% for TGC, and 20.58\%/28.98\% for Trans2vec. These results suggest that \sys generalizes much more effectively to previously unseen nodes and is better suited for real-world phishing detection, where malicious accounts and transactions emerge continuously and often cannot be observed during training.

\begin{table}[!t]
	\centering
	\small
	\caption{Comparing with TTAGN, TGC, and Trans2vec in detecting phishing transactions}	
	\begin{tabular}{llllllll}
		\hline
		{Metric}    & {TTAGN}~\cite{li2022ttagn} & {TGC}~\cite{li2023tgc} & {Trans2vec}~\cite{wu2020phishers} & {\sys} \\
		\hline
		{Precision} & 97.05/0.06                        & 98.73/0.01                    & 99.68/0.03                               & 99.25/0.02    \\
		{Recall}    & 71.28/0.17                        & 68.71/0.23                    & 39.45/0.12                               & 77.81/0.65    \\
		{FPR}       & 2.17/0.05                         & 0.88/0.00                     & 0.13/0.01                                & 0.59/0.02     \\
		{FNR}       & 28.72/0.17                        & 31.29/0.23                    & 60.55/0.12                               & 22.19/ 0.65   \\
		{F1}        & 82.19/0.09                        & 81.03/0.17                    & 56.53/0.12                               & 87.23/0.40    \\
		{BAC}       & 84.56/0.06                        & 83.92/0.12                    & 69.66/0.05                               & {88.61/0.32 } \\
		{AUC}       & 96.34/0.01                        & 98.10/0.02                    & 82.14/0.95                               & 98.43/0.04    \\
		\hline
	\end{tabular}	
	\label{tab:comparing_transactions}
\end{table}

\begin{table}[!t]
	\centering
	\small
	\caption{Comparing with TTAGN, TGC, and Trans2vec in detecting phishing accounts}	
	\begin{tabular}{llllll}
		\hline
		{Metric}    & {TTAGN}~\cite{li2022ttagn} & {TGC}~\cite{li2023tgc} & {Trans2vec}~\cite{wu2020phishers} & {\sys} \\
		\hline
		{Precision} & 89.44/0.60                        & 91.59/0.75                    & 83.17/2.20                               & 92.24/0.78    \\
		{Recall}    & 94.07/0.96                        & 96.13/1.40                    & 89.61/0.96                               & 96.24/0.94    \\
		{FPR}       & 11.15/1.15                        & 8.89/1.21                     & 18.11/2.27                               & 8.10/0.88     \\
		{FNR}       & 5.93/0.96                         & 3.87/1.40                     & 10.39/0.96                               & 3.76/0.94     \\
		{F1}        & 91.69/0.57                        & 93.80/0.91                    & 86.26/1.48                               & 94.19/0.49    \\
		{BAC}       & 91.46/0.79                        & 93.62/1.08                    & 85.75/1.35                               & {94.07/0.60}  \\
		{AUC}       & 96.20/0.46                        & 97.79/0.43                    & 91.95/1.40                               & 98.03/0.38    \\
		\hline
	\end{tabular}	
	\label{tab:comparing_accounts}
\end{table}

\subsection{Evaluation in Real-World Deployment}

Based on the above results, we further deployed a prototype of \sys using 128-dimensional representations and 20 sampled neighbors for real-world phishing detection on Ethereum from May 1, 2023 to July 31, 2024. During this period, \sys identified 2,153 phishing addresses involved in 73,832 transactions, demonstrating its ability to operate continuously on a large-scale dynamic transaction graph. More importantly, 1,803 of these addresses were detected before any public reports and were later confirmed through victim disclosures and phishing website analysis, showing that \sys can provide genuinely proactive early warnings rather than merely rediscovering already known threats. In addition, 196 more addresses were flagged as suspicious and subsequently validated by risk auditors. With only 47 phishing addresses missed, corresponding to an overall FNR of 2.14\%, \sys helped preempt potential losses of \$2,039.1M out of an estimated \$2,813.7M during the deployment period. These results highlight both the practical effectiveness of \sys and the persistent severity of phishing threats in the Ethereum ecosystem.

\begin{figure}[!t]
	\centering
	\subfloat[]{\includegraphics[width = 0.45\linewidth]{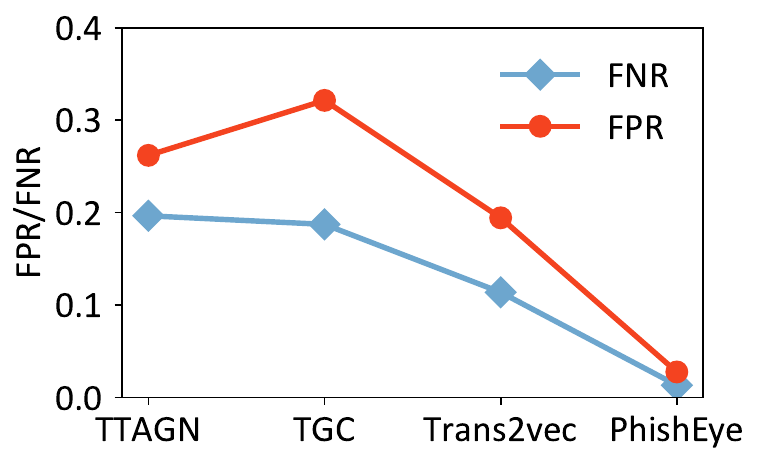}} \hfil
	\subfloat[]{\includegraphics[width = 0.45\linewidth]{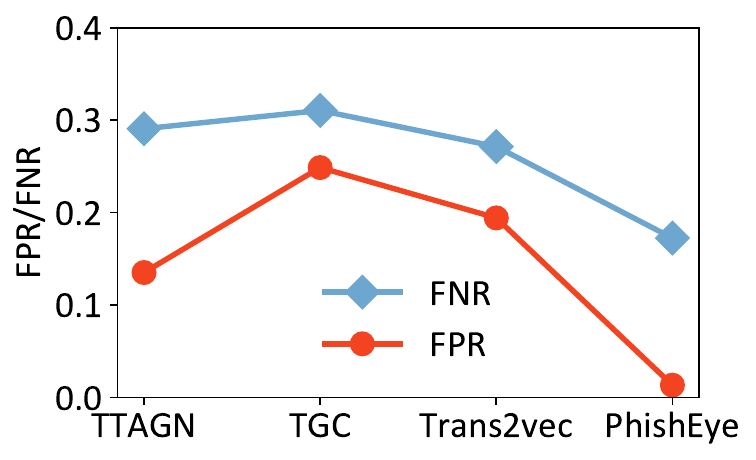}}	
	\caption{FNR/FPR of different methods in detecting real-world phishing accounts (a) and transactions (b)}
	\label{fig:real_error}	
\end{figure}

Figure~\ref{fig:real_error} further compares \sys with existing methods in this real-world setting. Overall, \sys achieves the most favorable trade-off between missed detections and false alarms for both phishing accounts and phishing transactions. In account detection, \sys reduces the error to 1.31\% FNR and 2.75\% FPR, substantially below the error ranges of TTAGN, TGC, and Trans2vec, whose false negative rates remain between 11.37\% and 19.65\% and false positive rates range from 19.43\% to 32.14\%. A similar trend is observed in transaction detection, where \sys maintains a very low FPR of 1.32\% while reducing the FNR to 17.24\%, clearly outperforming all baselines. These findings show that \sys is not only accurate in offline evaluation, but also robust and deployable in practice, where both timely detection and low false-alarm rates are critical for security operations.

\subsubsection{Inference Latency and Overhead}
We also evaluated the system overhead and processing efficiency of \sys on a device equipped with an Intel(R) Xeon(R) Gold 6133 CPU @ 2.50GHz and an NVIDIA GeForce RTX 4090 GPU. During the training phase, \sys consumed 21.10 GB of GPU memory and required approximately 392 minutes to complete 30 training epochs. In the inference phase, it used 6.5 GB of GPU memory, achieving an average delay of 2.16 milliseconds per address and 2.92 milliseconds per transaction. These results indicate that \sys has low system overhead and is capable of efficient large-scale inference, making it suitable for real-time blockchain phishing detection.

To further assess the practicality of \sys, we measured its end-to-end runtime overhead across all processing stages. The full pipeline consists of transaction data retrieval, temporal graph construction, node and edge feature extraction, and model inference. For a batch of 10,000 raw Ethereum transactions, data retrieval and preprocessing via Web3 interfaces take approximately 1.04 seconds, graph construction takes 0.53 seconds, and feature extraction, covering all 18 node features (T1-T9, N1-N9) and edge attributes, requires 1.82 seconds. Combined with the inference time, \sys completes the entire processing pipeline in under 3.54 seconds per batch, exhibiting the practical applicability for timely phishing detection on blockchain platforms.

\subsubsection{Case Study}
Table~\ref{tab:case_details} presents the phishing activities of three representative phishing cases identified by \sys.

\begin{table}[!t]
	\centering
	\footnotesize
	\caption{Distribution of fund flow destination. The detailed addresses are given in Appendix~\ref{app:addr}}	
	\begin{tabular}{lccr}
		\hline
		{Victim addr} & {Amount/ETH} & {Date} & {\# Transfer} \\ 		
		\hline
		\multicolumn{4}{l}{Case 1, Phishing addr: 0x69..55}                               \\
		0x99..88             & 9,257.8             & Sep-6-2023    & 2                    \\
		0x3f..ad             & 5,532.5             & Sep-6-2023    & 41                   \\ \hline
		\multicolumn{4}{l}{Case 2, Phishing addr: 0x1d..df}                               \\
		0xef..6b             & 258.3               & May-20-2023   & 161                  \\
		0x00..00             & 37.8                & May-18-2023   & 16                   \\
		0xba..05             & 32.0                & May-20-2023   & 12                   \\
		0x00..ac             & 13.6                & Jul-26-2023   & 13                   \\
		0x27..b6             & 12.2                & Apr-29-2023   & 6                    \\
		0xb8..7f             & 11.0                & May-20-2023   & 4                    \\ \hline
		\multicolumn{4}{l}{Case 3, Phishing addr: 0x9f..26}                               \\
		0x3f..ad             & 378.8               & Dec-26-2023   & 817                  \\
		0x00..00             & 281.6               & Jan-1-2024    & 607                  \\
		0x5c..c5             & 277.7               & Jan-1-2024    & 121                  \\
		0xef..6b             & 252.9               & Jun-6-2024    & 298                  \\
		0xxf..67             & 231.5               & Jan-1-2024    & 410                  \\
		0x11..82             & 220.3               & Dec-26-2023   & 293                  \\
		0x00..ac             & 205.0               & Jul-7-2023    & 12                   \\ 		
		\hline
	\end{tabular}
	\label{tab:case_details}
\end{table}

\textbf{Case 1-0x69..55.}
In a notable Ethereum phishing case, two addresses, i.e., 0x99..88 and 0x3f..ad, fell victim to an attack on September 6, 2023.
The phishing address 0x69..55 detected by \sys, successfully exfiltrated assets totaling {14,790.3 ETH (\$24M)} using {43} transfer events.
This included 4,851 units of Rocket Pool ETH and 9,579 units of Lido Staked ETH. The attack mechanism involved a phishing link, which tricked the victims into authorizing a malicious \texttt{transferFrom} function.
The funds were dispersed through various channels to conceal their origin and complicate tracking efforts.
The attacker quickly moved stolen assets to multiple addresses or through cryptocurrency exchanges to launder and extract the funds, making recovery challenging.
This enabled the attacker to divert the assets to their address, underscoring the acute vulnerabilities present in DeFi space. The incident accentuates the necessity for rigorous security measures and user vigilance in cryptocurrency transactions, especially when dealing with substantial assets.
The phishing funds were quickly moved to multiple addresses or through cryptocurrency exchanges to launder and extract the funds, making recovery challenging.

\textbf{Case 2-0x1d..df.}
Phishing address 0x1d..df identified by \sys orchestrated multi-chain phishing attacks, stealing approximately {364.9 ETH (\$0.6M)} from May to July 2023.
This operation victimized at least {6 addresses} (0xef..6b, 0x00..00, 0xba..05, 0x00..ac, 0x27..b6, 0xb8..7f) with losses exceeding {10 ETH} each in {212} transfer events
One notable victim alone lost {\$0.48M}.
After our further investigation and tracking, we found the attack address is linked to Inferno Drainer~\cite{Inferno_Drainer}, a notorious phishing entity.
Inferno Drainer's extensive phishing campaign impacted nearly 4,888 victims and amassed around \$5.9M across various chains. They created over 9500 phishing sites targeting more than 220 popular brands and projects, employing diverse tactics to deceive users and expropriate assets.
Analysis of on-chain fund movements revealed significant ETH distributions among several addresses, highlighting the elaborate and extensive nature of Inferno Drainer's phishing operations in the crypto space.

\textbf{Case 3-0x9f..26.}
Phishing address 0x9f..26 executed a major operation, successfully extracting {1,847.8 ETH} in {2,558} transfers, valued at around {\$3.6M}, between July 2023 and January 2024. The attack impacted at least {7 addresses}, each with over {200 ETH}, showcasing the extent and sophistication of the phishing strategy employed. This case underscores the scale of the operation and the significant impact it had on its victims.
Under our further verification and investigation, we found the group's phishing strategy is operated by creating phishing websites and executing social engineering attacks.
For instance, they impersonated media outlets and conducted interviews that ultimately led to Discord token theft. The stolen funds were distributed among various addresses.

\begin{figure}[!t]
	\centering
	\subfloat[Fund flow]{\includegraphics[width = 0.38\linewidth]{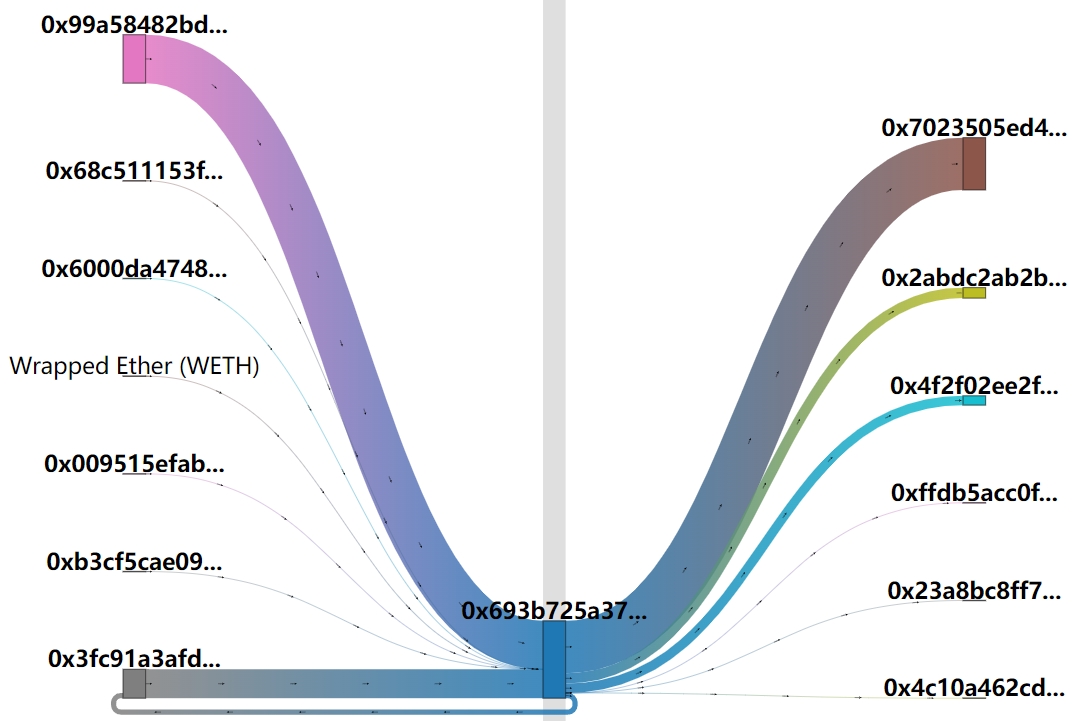}} \hspace{1mm}
	\subfloat[Inflow transfers in ETH currency ]{\includegraphics[width = 0.58\linewidth]{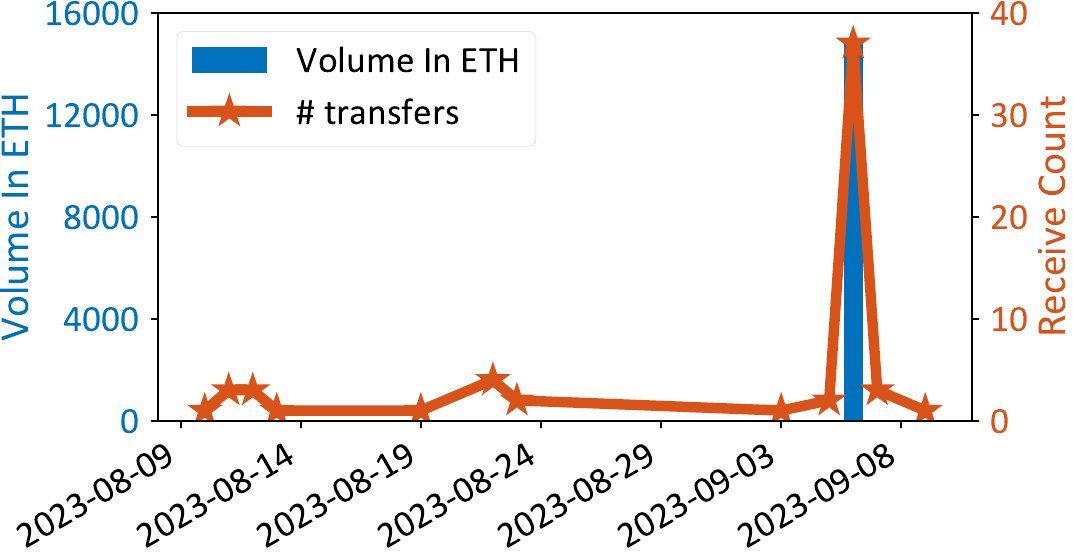}}	
	\caption{Case 1  with the phishing address, 0x69..55: fund flow (a) and inflow transfers in ETH currency (b)}
	\label{fig:case1}
\end{figure}

\begin{figure}[!t]
	\centering
	\subfloat[Fund flow]{\includegraphics[width = 0.38\linewidth]{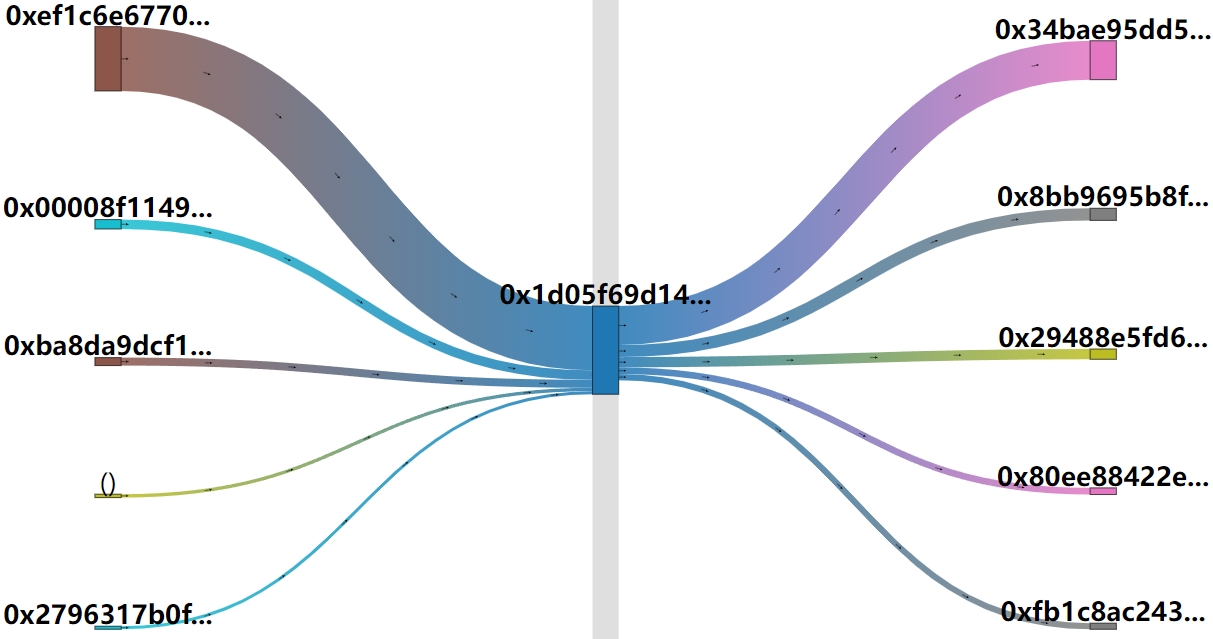}} \hspace{1mm}
	\subfloat[Inflow transfers in ETH currency ]{\includegraphics[width = 0.58\linewidth]{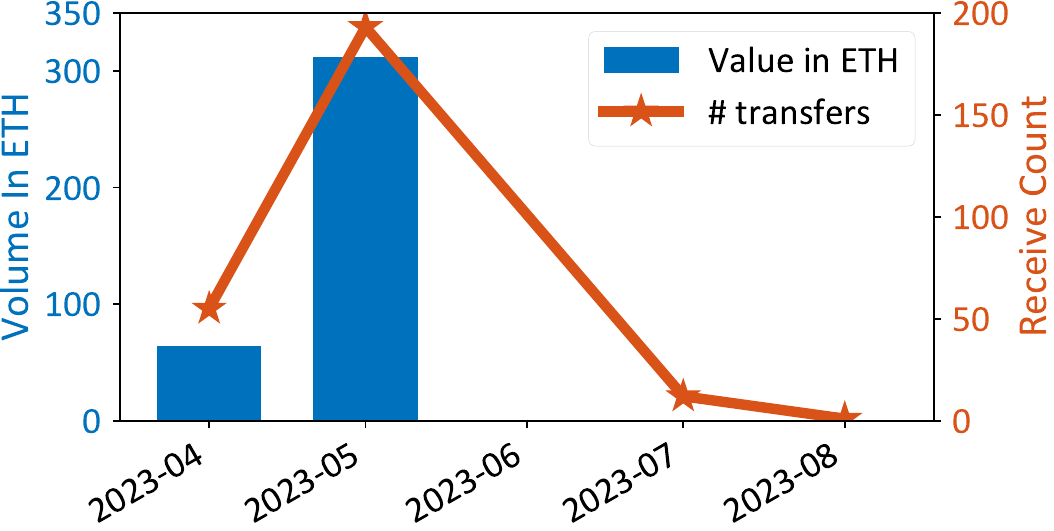}}	
	\caption{Case 2 with the phishing address, 0x1d..df: fund flow (a) and inflow transfers in ETH currency (b)}
	\label{fig:case2}	
\end{figure}

\begin{figure}[!t]
	\centering
	\subfloat[Fund flow]{\includegraphics[width = 0.38\linewidth]{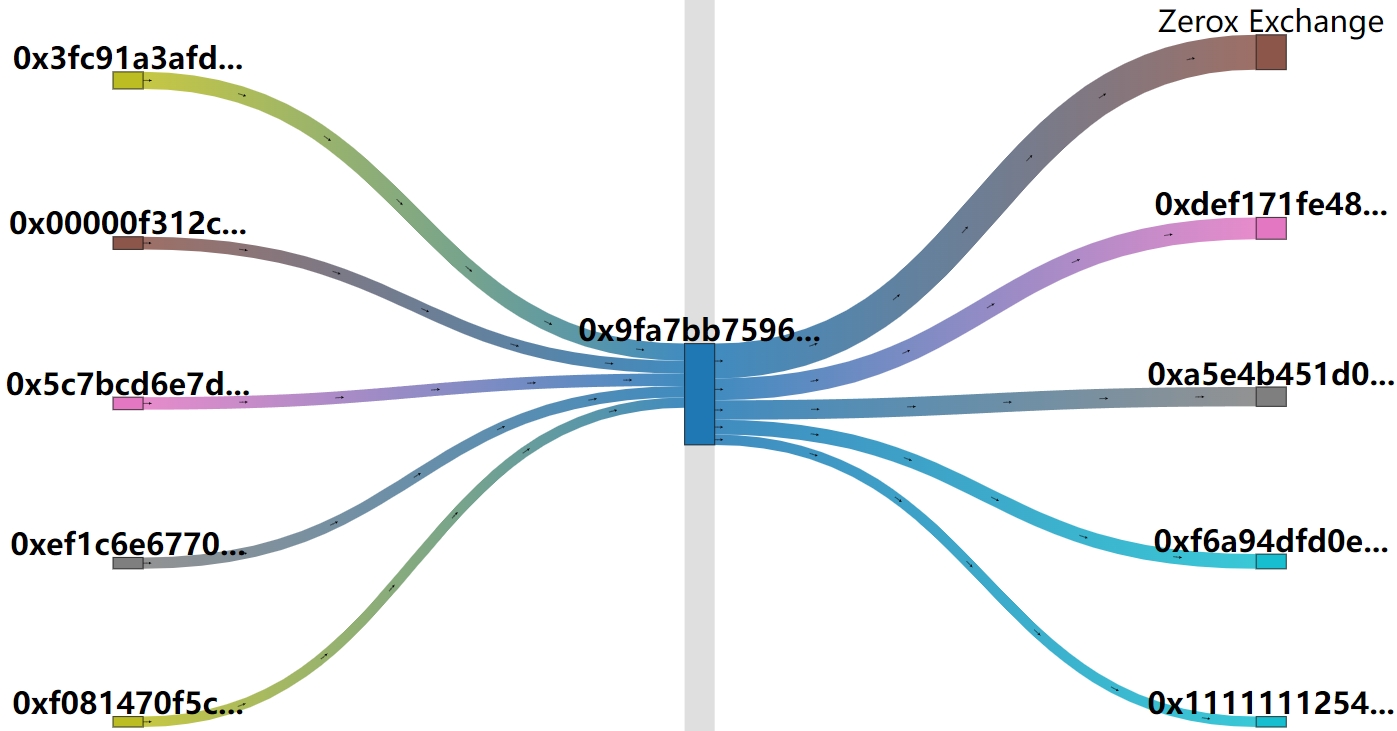}} \hspace{1mm}
	\subfloat[Inflow transfers in ETH currency ]{\includegraphics[width = 0.58\linewidth]{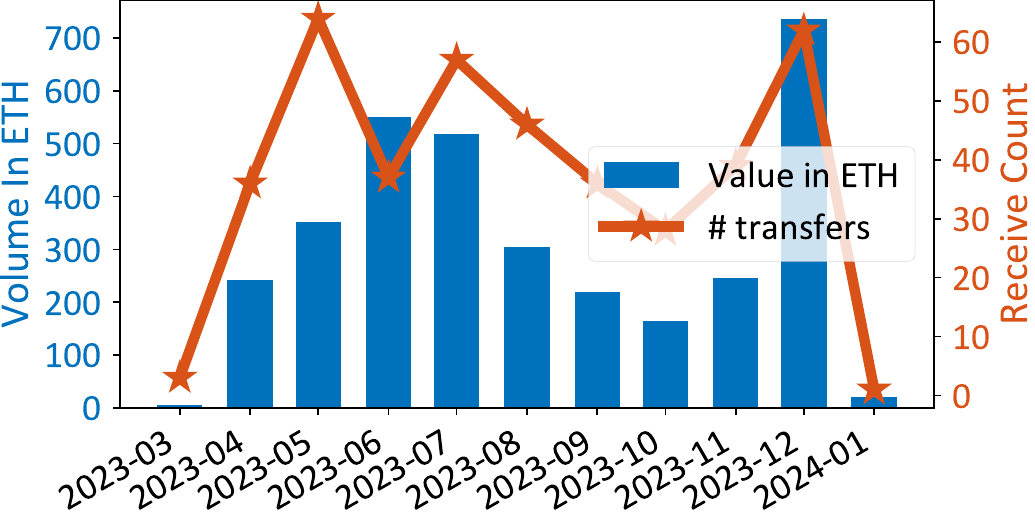}}\
	\caption{Case 3 with the phishing address, 0x9f..26: fund flow (a) and inflow transfers in ETH currency (b)}
	\label{fig:case3}	
\end{figure}

\section{Related Work}

Previous research on blockchain phishing consists of both non-graph and graph-based approaches, leveraging signals from promotions, website content, and social or transactional contexts.
Non-graph-based approaches typically rely on statistical features, such as transaction frequency and neighbor behavior, and employ supervised models like CNNs or LSTMs~\cite{wen2022hide,chen2020phishing}. However, these methods overlook the inherent relational structure within the transaction network.

Roy et al.~\cite{roy2023demystifying} delved into the sphere of NFT promotion phishing scams, developing sophisticated ML models capable of identifying fraudulent NFT projects and phishing websites.
 Kim et al.~\cite{kim2023drainclog} employed GNNs to analyze the transaction and social contexts, implementing an automated system to detect NFT phishers.
These developments underscore the increasing complexity and necessity for advanced methodologies in combating phishing in the evolving landscape of blockchain and NFTs.
TTAGN~\cite{li2022ttagn} detected phishing transactions using LSTM to embed temporal characteristics of transactions and using graph convolutional model to extract structural characteristics.
TxPhishScope~\cite{he2023txphishscope} detected phishing websites and extracted phishing accounts by scanning suspicious websites and triggering transactions.
Besides, token-related scams were widely explored. Chen et al.~\cite{chen2020traveling} introduced a transaction graph-based feature extraction for scam token analysis. Xia et al.~\cite{xia2021trade} proposed extracting diverse features, including time-series and transaction data, to train classifiers for detecting rugpull tokens.

A high-level comparison of \sys is detailed in Table~\ref{tab:comparison_sys}. Unlike existing methods, \sys relies on temporal encoding and temporal graph learning on heterogeneous graphs. TTAGN~\cite{li2022ttagn} combines static GNN with LSTM, partially supports temporal encoding, and relies on supervised learning. In contrast, \sys fully integrates temporal encoding and self-supervised learning, addressing the challenge of accurately capturing the evolving nature of phishing activities. Trans2vec~\cite{wu2020phishers} and TGC~\cite{li2023tgc} use random walk methods, lacking comprehensive temporal and heterogeneous graph handling capabilities. \sys advances these by employing sophisticated temporal graph learning techniques.
\sys is particularly robust in detecting both transactions and accounts, designed for effective operation in real-world environments, and capable of handling heterogeneous graphs. Other methods, like those by Chen et al.~\cite{chen2020phishing} and Wen et al.~\cite{wen2022hide}, use static features and fail to address temporal dynamics and self-supervised learning comprehensively. \sys uniquely addresses these gaps by providing a robust solution for detecting phishing activities in dynamic and complex blockchain environments. Additionally, the real-world evaluation conducted demonstrates \sys's effectiveness in practical scenarios, further highlighting its superiority.
\begin{table}[!t]
	\centering
	\footnotesize
	\caption{High-level comparison of existing methods (\Circle: not fulfilling the specified condition, \CIRCLE: fully fulfilling, \LEFTcircle: partially fulfilling)}
	\begin{tabular}{p{42pt}p{85pt}p{3pt}p{3pt}p{3pt}p{3pt}p{6pt}p{6pt}}
		\hline
		{Method}                    & {Method/features}    & {TE} & {SSL} & {DT} & {DA} & {RWE} & {HEG} \\
		\hline
		TTAGN~\cite{li2022ttagn}            & Static GNN + LSTM             & \LEFTcircle & \Circle      & \CIRCLE     & \Circle     & \Circle      & \Circle      \\
		Trans2vec~\cite{wu2020phishers}     & RW + OC-SVM                   & \Circle     & \Circle      & \Circle     & \CIRCLE     & \Circle      & \Circle      \\
		TGC~\cite{li2023tgc}               & RW + GCN                    & \Circle     & \Circle      & \Circle     & \CIRCLE     & \Circle      & \Circle      \\
		Chen et al.~\cite{chen2020phishing} & Statistic features + LightGBM & \Circle     & \Circle      & \CIRCLE     & \Circle     & \Circle      & \Circle      \\
		Wen et al.~\cite{wen2022hide}       & Statistic features + GS       & \Circle     & \Circle      & \Circle     & \CIRCLE     & \Circle      & \Circle      \\ \hline
		{\sys}                     & Temporal graph learning     & \CIRCLE     & \CIRCLE      & \CIRCLE     & \CIRCLE     & \CIRCLE      & \CIRCLE      \\
		\hline
	\end{tabular}
	\label{tab:comparison_sys}
	\begin{tablenotes}
		\item TE: temporal encoding. SSL: self-supervised learning. DT: detecting transaction. DA: detecting account.
		RWE: real-world detection.
		HEG: heterogeneous graph.
		RW: random walk.  OC-SVM: one-class support vector machine. GCN: graph convolution network. GS: greedy selection.
	\end{tablenotes}
	\end{table}

\textbf{Phishing measurement.}  
A wide range of measurement studies in Web3 security have been conducted to explore specific threats and behaviors within blockchain ecosystems, including phishing attacks, fraudulent promotions, token manipulation, etc.
Yang et al.~\cite{yang2023trail} offered an in-depth analysis of NFT phishing attacks, categorizing them into distinct patterns, while Chen et al.~\cite{chen2020traveling} delved into the Ethereum ERC20 token ecosystem, examining its distribution and circulation. Lyu et al.~\cite{lyu2022empirical} focused on private transactions on Ethereum, assessing their impact on blockchain security. Roy et al.~\cite{roy2023demystifying} analyzed Twitter NFT promotions, unveiling a high incidence of scams and developing tools for fraud detection. Zheng et al.'s work on the EOSIO blockchain~\cite{zheng2022unravelling} investigated the structural and transactional aspects of token relationships. Das et al.~\cite{das2022understanding} discussed vulnerabilities within the NFT ecosystem and proposed security enhancements. Li et al.~\cite{li2023double} analyzed the flow of stolen funds on public blockchains, while Xu et al.~\cite{xu2019anatomy} examined pump-and-dump schemes in Telegram channels. Cernera et al.~\cite{cernera2023token} studied token and liquidity pool scams on Ethereum and BNB Smart Chain, contributing to a comprehensive understanding of security issues within evolving blockchain ecosystems.

\section{Limitation and Future Work}
\label{app:discussion}
In this section, we discuss the limitations of \sys and outline potential directions for future research improvements.

\textbf{Using off-chain information.}
We strive to maintain a robust phishing detection system using only on-chain data, but some limitations may exist. 
First, while excluding off-chain data can reduce complexity, it limits the model's ability to fully capture phishing patterns that extend beyond on-chain transactions. Many phishing websites and malicious activities remain unreported or hidden off-chain, making comprehensive tracking and detection challenging. Furthermore, phishing addresses are continually evolving, and without off-chain sources, updating the model with real-time data becomes difficult, impacting detection accuracy and timeliness.

Additionally, the dataset used for training \sys primarily focuses on phishing transactions and accounts, which improves the model's detection efficiency but reduces data diversity.
This targeted approach ensures the model effectively learns phishing-specific behaviors but may reduce its ability to generalize to other transaction patterns and account behaviors.
While we have taken great care to ensure the accuracy of the phishing labels in our dataset through rigorous cross-verification, there may be occasional labeling errors. Few phishing addresses could have been misclassified due to the dynamic and deceptive nature of phishing schemes, where new tactics and addresses frequently emerge. 

\textbf{Future Work.}
In the future, efforts may focus on expanding the dataset to include a broader range of Ethereum transactions, both phishing and non-phishing, to improve the model's generalizability across various transaction types. 
Additionally, it is necessary to test \sys on other blockchain networks, such as BTC chain and Binance Smart Chain, to evaluate its effectiveness in different environments.
Future research can integrate multi-modal data sources, including off-chain reports and social media signals, to enhance phishing detection across both on-chain and off-chain interactions. Furthermore, advanced anomaly detection techniques will be explored to identify evolving phishing tactics and ensure \sys remains effective against emerging threats, keeping it relevant and adaptive in real-world scenarios.

\section{Conclusion}

In this paper, we presented \sys, a temporal graph contrastive learning-based system for detecting cryptocurrency phishing activities on Ethereum blockchain. 
By modeling on-chain transaction data as an HTAMG and incorporating the custom-designed PhishTGL model, \sys effectively captures the temporal dynamics and heterogeneous transaction types inherent to phishing attacks. 
Unlike existing static and semi-dynamic approaches, \sys utilizes a self-supervised learning strategy, which allows it to learn robust node and edge representations without relying on extensive labeled data.
 Our comprehensive evaluations demonstrate significant performance gains over state-of-the-art methods, achieving superior F1 scores and AUC values for both phishing transaction and account detection. 
 Furthermore, a real-world deployment of \sys successfully identified 1,803 previously unknown phishing addresses, alerting an estimated over 2 billion USD in financial losses. 

\section*{Ethical Considerations}

This work was conducted with careful attention to the ethical and operational implications of phishing detection in blockchain ecosystems. Our study does not involve human-subject experiments, direct interaction with victims, intervention on user accounts, or any active engagement with attackers. All analyses are performed on publicly accessible Ethereum on-chain transaction records, which are pseudonymous by design. We do not attempt to deanonymize users, infer real-world identities, or combine on-chain data with external personal information for re-identification. The goal of this work is limited to identifying suspicious transaction patterns and phishing-controlled addresses from observable on-chain behavior.

We also considered the dual-use risk of releasing a detection system for adversarial activity. To reduce the possibility of misuse, the paper focuses on defensive methodology, aggregate evaluation, and security insights rather than operational details that could facilitate evasion or abuse. The released artifacts do not include any private keys, exploit scripts, or operational instructions that could enable phishing campaigns. In the real-world deployment described in this paper, detected phishing addresses were responsibly disclosed to anonymized security partners for further alerting and tracking, rather than being used for public accusation or disruptive action.

In addition, we acknowledge that blockchain security detection can produce false positives and should not be treated as standalone proof of malicious intent. For this reason, the system is intended to support risk screening and early warning, and suspicious cases are further validated through victim disclosures, phishing site analysis, and auditor review. We believe these safeguards are important for minimizing harm and supporting the responsible use of our findings. A detailed ethics discussion is provided in the appendix.

\section*{Open Science}
Our work complies with the Open Science Policy. We provide an anonymous artifact at \url{https://anonymous.4open.science/r/PhishEye-3CEE/}. The repository includes the implementation of PhishEye, together with the scripts, configuration files, and documentation needed to reproduce the main experimental results reported in the paper. Where applicable, we also provide the released benchmark resources and detailed instructions for training, evaluation, and deployment-related reproduction.

\bibliographystyle{ACM-Reference-Format}
\bibliography{citiation}

\clearpage

\appendix

\section{Extracted Node Features by \sys} 
\label{app:nodefeatures}
In detail, \sys extracts transactional activity features (T1-9) and network
structure features (N1-9) for each account address in the transaction graph.
\begin{inditemize}
	\item  \emph{T1-2. Account indegree and outdegree (AID, AOD)} captures the number of incoming and outgoing transactions to or from node \( v \) respectively, defined as $ aid_v = |{e(u,v,t_e)|,u \in N(v), e(u,v,t_e)\in \mathcal{E}}$, and
	$ aod_v = |{e(v,u,t_e)|,u \in N(v), e(u,v,t_e)\in \mathcal{E}}$,
	where $e(u,v,t_e)$ represents the edge from node $u$ to $v$ at time $t_e$.
	$N(v)$ are node $v$'s neighbors.
	AID and AOD reflect the node's receiving and sending activity, respectively.

	\item \emph{T3. Account total degree (ATD)} captures the total number of transactions of an account, defined as $ td_v = aid_v +  aod_v$.
	It encapsulates the overall transactional engagement of an account, considering its receiving and sending activities.

	\item \emph{T4-5. Inbound and outbound value flow (IVF, OVF)} quantifies the total received and sent fund of node $v$, i.e., $ ivf_v = \sum_{u \in In(N(v))} a_{uv} $ and $ ovf_v = \sum_{u \in Out(N(v))} a_{vu} $,  where \( a_{uv} \) is the transaction amount from node \( u \) to \( v \).
	They measure a node's economic interactions within the transaction network, encapsulating both the inflow and outflow of transaction values and representing the node's economic activity in incoming and outgoing Ether transfers.

	\item \emph{T6. Total value exchange (TVE)}  represents the total exchanged value of a node, defined as $ tvf_v = tvf_v+ ovf_v$.
	TVE provides a comprehensive measure of an account's total engagement in transfer value, representing a node's economic activity and influence.

	\item \emph{T7. Degree temporal density (DTD)}  captures the essence of how densely transaction connections are distributed over time for each node, i.e., $DTD_v = D_v/(t_{vn}- t_{v1})$, where \( D_v \) is the total transaction degree of node \( v \), \( t_{vn} \) and \( t_{v1} \) are the timestamps of the last transaction and the first transaction. It quantifies the frequency of a node's transactions over time, reflecting its temporal transactional activity.

	\item \emph{T8. Contract interaction ratio (CIR)} measures the proportion of transactions that involve smart contract interactions relative to the total number of transactions for each node.
	CIR represents how actively a node engages with smart contracts in a transaction graph.

	\item \emph{T9. Peak neighbor transaction degree (PNTD)} captures the essence of the maximum transactional engagement a node has with any of its neighbors, defined as $pntd_v = \max \left( \{ |ATD_u| \, : \, u \in N(v) \} \right)$.
	PNTD emphasizes the node's most regular interaction and the primary transactional relationship from a graphical perspective.

	\item  \emph{N1. Node connectivity degree (NCD)} measures how many different accounts each node has transacted with, i.e., $ ncd_v= N(v)|$. It represents the node's interaction level, highlighting its relative importance and influence.
	A high NCD indicates a node that is central to numerous nodes,
	playing a significant role in economic activity.

	\item  \emph{N2-3. Indegree and outdegree centrality (IC, OC)} represents the node importance as a recipient and sender in a transaction network based on the transactional incoming and outgoing connections.
	INC measures the proportion of incoming edges a node has, compared to rest nodes in the network, defined as $aid_v/(|\mathcal{V}|-1)$, indicating the an account's activeness is a recipient in transactions. OUC calculates the proportion of outgoing edges, defined as $aod_v/(|\mathcal{V}|-1)$, reflecting the activeness of an account that initiates transactions.

	\item \emph{N4. Degree centrality (DC)} represents a node's overall importance based on its total transactional connections, i.e., $(aid_v+ aod_v)/(|\mathcal{V}|-1)$.
	It measures the number of transactions a node has relative to the rest nodes,
	and highlights nodes' activeness.

	\item \emph{N5. Average neighbor degree centrality (ANDC)} measures the average degree centrality for a node neighbor, i.e.,  $(\sum_{u\in N(v)dc_u})/|N(v)|$. ANDC reflects the average importance of node neighbors.

	\item \emph{N6. Pagerank (PR)} measures the node's relative importance based on the structure of incoming links and the rank of the linking nodes, defined as $pr_v = (1-d)/|\mathcal{V}| + d\sum_{u\in N(v)}^{} \frac{pr_u}{aod_v}$, where $d$ is the damping factor.

	\item \emph{N7. Square clustering coefficient (SCC)} measures the prevalence of squares (cycles of length four) that \( v \) is a part of.
	SCC represents higher-order connectivity around a node, indicating the existence of complex transaction patterns.

	\item \emph{N8. Maximal network reach (MNR)} measures the maximum distance of \( v \) to any other node defined as: $ mnr_v = \max (d(v,u) :u \in \mathcal{V})$,
	where \( d(v, u) \) is the shortest path distance between nodes \( v \) and \( u \).
	MNR reflects the transactional relative remoteness or accessibility of a node.

	\item \emph{N9. Unique transactional pathways (UTP)} measures the number of unique paths that start from a node that reach other the node's non-$1_{st}$ order neighbors without revisiting any node.
	UTP indicates the node's capability to establish diverse transactional connections.
\end{inditemize}

\section{Address Details}
\label{app:addr}
To better track the phishing cases in our work, we also present the detailed addresses in Table~\ref{fig:addrs}.

\begin{table}[!h]
	\centering
	\scriptsize
	\caption{Address details}
	\resizebox{.8\linewidth}{!}{\begin{tabular}{lr}
		\hline
		Short address & Full address                               \\ 				
		\hline
		0x00..ac      & 0x0000000000a39bb272e79075ade125fd351887ac \\
		0x3f..ad      & 0x3fc91a3afd70395cd496c647d5a6cc9d4b2b7fad \\
		0x00..00      & 0x00000f312c54d0dd25888ee9cdc3dee988700000 \\
		0x5c..c5      & 0x5c7bcd6e7de5423a257d81b442095a1a6ced35c5 \\
		0xef..6b      & 0xef1c6e67703c7bd7107eed8303fbe6ec2554bf6b \\
		0xf0..67      & 0xf081470f5c6fbccf48cc4e5b82dd926409dcdd67 \\
		0x11..82      & 0x1111111254eeb25477b68fb85ed929f73a960582 \\
		0x00..00      & 0x00008f1149168c1d2fa1eba1ad3e9cd644510000 \\
		0xba..05      & 0xba8da9dcf11b50b03fd5284f164ef5cdef910705 \\
		0x00..ac      & 0x0000000000a39bb272e79075ade125fd351887ac \\
		0x27..b6      & 0x2796317b0ff8538f253012862c06787adfb8ceb6 \\
		0xb8..7f      & 0xb8901acb165ed027e32754e0ffe830802919727f \\
		0x9f..26      & 0x9fa7bb759641fcd37fe4ae41f725e0f653f2c726 \\
		0x99..8f      & 0x99a58482bd75cbab83b27ec03ca68ff489b5788f \\
		0x1d..df      & 0x1d05f69d14519a1c93007b468e30d5cb1f1658df \\
		0x69..55      & 0x693b725a375f599f0b6efa0d910e749e1eec1555 \\
		\hline
	\end{tabular}}
	\label{fig:addrs}
\end{table}

\section{Interfaces and Emitted Events of Ethereum Tokens}
\label{app:tokenfuns}
Table~\ref{tab:erc-functions} presents the detailed Ehereum tokens' main interface and emitted events.

\begin{table*}[!t]
	\scriptsize
	\centering
	\caption{Ethereum tokens' main interface and emitted events, including approval and transfer. \texttt{Approval} interface grants permission for a \texttt{to} smart contract to transfer a specified \texttt{value} or \texttt{tokenid} NFT on behalf of the \texttt{owner}.
	\texttt{Transfer} interface moves \texttt{value} FTs or \texttt{tokenid} NFT from account \texttt{from} account to \texttt{to}.
	Approval and transfer events are emitted notifications that log and broadcast approvals and token transfers to external observers.}
	\label{tab:erc-functions}
	\resizebox{\linewidth}{!}{\begin{tabular}{p{26mm}|ll|llll}
		\hline
			\textbf{Token standard}          & \textbf{Approval interface}             & \textbf{Transfer interface}                      & \textbf{Approval events}                   & \textbf{Transfer events}                           \\ 		
			\hline
			\textbf{ERC-20 (FT)}             & \texttt{Approve\texttt{(to,value)}}     & \texttt{transferFrom(from,to,value)}             & \texttt{Approval(owner,to,value)}          & \texttt{Transfer(from,to,value)}                   \\ \hline
			\textbf{ERC-721 (NFT)}           & \texttt{Approveal(to,tokenid)}          & \texttt{safeTransferFrom(from,to,tokenid)}       & \texttt{Approval(owner,to,tokenid)}
			                                 & \texttt{Transfer(from,to,tokenid)}                                                                                                                                                           \\ \hline
			\textbf{ERC-1155  (Multi-token)} & \texttt{setApprovalForAll(to,approved)} & \texttt{safeTransferFrom(from,to,id,value,data)} & \texttt{ApprovalForAll(owner,to,approved)} & \texttt{TransferSingle(operator,from,to,id,value)}
			\\ 				
				\hline
		\end{tabular}}
\end{table*}

\section{Phishing Fund Flow Analyzing}
In this section, we analyze fund flows of detected phishing addresses to understand the post-phishing behaviors.

\subsection{Tracking Phishing Funds}
We begin by extracting Ethereum transaction data  (S1), including sender and recipient addresses, token type, value, timestamp, and transaction hash, where timestamps provide chronological context and hashes ensure uniqueness. Next, we construct a fund flow graph (S2), representing accounts as nodes and transactions as timestamped, valued edges. This graph is continuously updated to capture evolving relationships and temporal patterns critical for analyzing complex fund movements. For efficiency, we apply strict termination rules (S3), halting analysis when transactions involve DEXs, CEXs, mixers, or bridges; reach a recursion depth over 10; involve super nodes with over 10k transactions; or show inactivity for over 720 days. We also filter out low-impact transactions, illiquid tokens, and chains with sparse activity. In S4, we track follow-up transactions involving the same tokens or addresses identified in S1, recursively analyzing recipients and recording the exact token amounts to maintain accuracy and prevent double-counting.

\subsection{Destination of Phishing Fund}
\begin{table}[!b]
	\centering
	\renewcommand{\arraystretch}{0.87}
	\caption{Destination of phishing stolen funds}
	\resizebox{\linewidth}{!}{\begin{tabular}{lrrrrr}
		\hline
		\textbf{Destination}   & \textbf{Total}      & \textbf{Ratio} & \textbf{Popular} & \textbf{Amount} & \textbf{Proportion} \\ 		
		\hline
		\multirow{6}{*}{CEX}                 & \multirow{6}{*}{\$174.0M}  & \multirow{6}{*}{31.9\%}   & Binance                       & \$103.6M        & 19.0\%              \\
		                                     &                            &                           & Coinbase                      & \$45.1M         & 8.3\%               \\
		                                     &                            &                           & Whalesheaven                  & \$6.1M          & 1.1\%               \\
		                                     &                            &                           & SimpleSwap                    & \$5.3M          & 1.0\%               \\
		                                     &                            &                           & Kraken                        & \$2.4M          & 0.4\%               \\
		                                     &                            &                           & Others                        & \$11.6M         & 2.1\%               \\ \hline
		\multirow{6}{*}{Mixer}               & \multirow{6}{*}{\$118.8M } & \multirow{6}{*}{ 21.8\%}  & Tornado Cash                  & \$85.7M         & 15.7\%              \\
		                                     &                            &                           & Sinbad                        & \$16.3M         & 3.0\%               \\
		                                     &                            &                           & FixedFloat                    & \$5.3M          & 1.0\%               \\
		                                     &                            &                           & MixEth                        & \$2.7M          & 0.5\%               \\
		                                     &                            &                           & Aztec                         & \$2.4M          & 0.4\%               \\
		                                     &                            &                           & Others                        & \$6.4M          & 1.2\%               \\ \hline
		\multirow{4}{*}{Cross-chain bridges} & \multirow{4}{*}{\$56.7M }  & \multirow{4}{*}{10.4\%}   & Avalanche                     & \$23.6M         & 4.3\%               \\
		                                     &                            &                           & Polygon PoS                   & \$9.1M          & 1.7\%               \\
		                                     &                            &                           & Wormhole                      & \$10.9M         & 2.0\%               \\
		                                     &                            &                           & Others                        & \$13.1M         & 2.4\%               \\ \hline
		\multirow{4}{*}{DEX}                 & \multirow{4}{*}{\$34.9M }  & \multirow{4}{*}{6.4\%}    & Uniswap                       & \$22.0          & 4.0\%               \\
		                                     &                            &                           & sushi                         & \$2.4M          & 0.4\%               \\
		                                     &                            &                           & 1inch                         & \$2.0M          & 0.4\%               \\
		                                     &                            &                           & Others                        & \$8.6M          & 1.6\%               \\ \hline
		In account balance                   & \$66.6M                    & 12.2\%                    & n.a                           & \$66.6M         & 12.2\%              \\ \hline
		Other methods                        & \$94.6M                    & 17.3\%                    & n.a                           & \$94.6M         & 17.3\%              \\ 		
		\hline
	\end{tabular}}
	\label{fig:flow_funds}
\end{table}

We use the phishing addresses detected during the period from October 1 to December 31, 2023, to analyze fund flow patterns, as this timeframe provides a representative set of confirmed cases with complete transaction histories for post-phishing behavior analysis.
As shown in Table~\ref{fig:flow_funds}, the flow distribution of stolen phishing funds provides a detailed picture of the channels used for asset management post-theft. CEXs are the primary destination, receiving {\$174.0M (31.9\%)}, with Binance alone accounting for {\$103.6M (19.0\%)}. Mixers, used for obfuscating transaction histories, received {\$118.8M (21.8\%)}, with Tornado Cash~\cite{tornado} being the most popular, receiving {\$85.7M (15.7\%)}. Cross-chain bridges, enabling asset transfers between different blockchains, accounted for {\$56.7M (10.4\%)}, with Avalanche receiving the highest amount of {\$23.6M (4.3\%)}. DEXs saw {\$34.9M (6.4\%)}, led by Uniswap with{ \$22.0M (4.0\%)}. Additionally, {\$66.6M (12.2\%)} remained in account balances, and other methods accounted for {\$94.6M (17.3\%)} of the stolen funds. 

CEXs are a primary destination for phishing funds, valued at \$174.0M (31.9\%). Binance, the most popular CEX, received \$103.6M (19.0\%). This preference for Binance could be attributed to its previous relatively relaxed regulation requirements~\cite{kyc}. Other significant exchanges used include Coinbase with \$45.1M (8.3\%) and Whalesheaven with \$6.1M (1.1\%). This trend underscores the role of CEXs in the laundering and conversion of illicitly obtained assets.

Mixers~\cite{mixers} play a pivotal role in phishing operations due to their ability to anonymize transactions. A substantial sum of \$118.8M, accounting for 21.8\% of the total phishing funds, was channeled through these mixers. Tornado Cash, in particular, processed \$85.7M, representing 15.7\% of this amount. Mixers are favored by attackers for their effectiveness in severing the traceability link between the original source and the final destination of the stolen funds, making tracking and recovery efforts significantly more challenging.

Cross-chain bridges~\cite{Bridge}, e.g., Avalanche, emerged as significant channels in phishing operations, accounting for \$56.7M (10.4\%) phishing funds. The utilization of cross-chain bridges in these phishing schemes is indicative of attackers' strategies to transfer assets across different blockchain networks, further complicating the tracking and recovery process. Avalanche bridge, e.g., was used to transfer \$23.6M, highlighting the main role in these cybercrimes.

DEXs, e.g., Uniswap and SushiSwap, play a significant role in phishing fund transfers. We found that DEXs facilitated \$34.9M, about 6.4\% of the total, with Uniswap accounting for \$22.0M. These platforms offer anonymity and ease of access, making them attractive for illicit transfers. Additionally, complex contract invoking, NFT exchanges, and conversions to other cryptocurrencies or assets on less-known DEXs accounted for \$94.6M, or 17.3\% of the total, highlighting diverse techniques used by attackers to obscure the origins and destinations of stolen assets.

\end{document}